\documentclass[10pt,journal,compsoc]{IEEEtran} 

\usepackage{cite} 
\usepackage{amsmath,amsfonts}

\usepackage{amssymb} 

\usepackage{algorithm}
\usepackage{algorithmicx,algpseudocode}
\usepackage{array}
\usepackage[caption=false,font=footnotesize]{subfig} 

\usepackage{textcomp}
\usepackage{stfloats} 
\usepackage{url}
\usepackage{verbatim}
\usepackage{graphicx}
\usepackage{xcolor} 

\usepackage{forest}
\usepackage{multirow} 
\usepackage{balance} 

\usepackage{rotating} 
\usepackage{booktabs}
\usepackage[colorlinks=true, linkcolor=black, urlcolor=black, citecolor=black]{hyperref}
\usepackage{listings}

\lstdefinestyle{calltree}{
  basicstyle=\footnotesize\ttfamily\color{black},
  frame=single,
  framerule=0.5pt,
  rulecolor=\color{black!60},
  backgroundcolor=\color{white},
  breaklines=true
}

\def\BibTeX{{\rm B\kern-.05em{\sc i\kern-.025em b}\kern-.08em
    T\kern-.1667em\lower.7ex\hbox{E}\kern-.125emX}}

\begin{document}

\title{TraDE: Network and Traffic-aware Adaptive Scheduling for Microservices Under Dynamics}

\author{Ming Chen, Muhammed Tawfiqul Islam, Maria Rodriguez Read, and Rajkumar Buyya
    \thanks{Manuscript submitted October, 2024.}
     \thanks{Ming Chen, Muhammed Tawfiqul Islam, Maria Rodriguez Read, and Rajkumar Buyya are with the Quantum Cloud Computing and Distributed Systems (qCLOUDS) Laboratory, School of Computing and Information Systems, The University of Melbourne, Australia. (e-mail: mingc4@student.unimelb.edu.au; {tawfiqul.islam, maria.read, rbuyya}@unimelb.edu.au)}
}


\maketitle

\begin{abstract}
The transition from monolithic architecture to microservices has enhanced flexibility in application design and its scalable execution. This approach typically uses a computing cluster managed by a container orchestration platform to deploy microservices. However, this shift introduces significant challenges, particularly in the efficient scheduling of containerized services. These challenges are compounded by unpredictable scenarios such as dynamic incoming workloads with various execution traffic and variable communication delays among cluster nodes. Existing works often overlook the real-time traffic impacts of dynamic requests on running microservices, as well as the varied communication delays across cluster nodes. Consequently, even optimally deployed microservices could suffer from significant performance degradation over time. To address these issues, we propose a network and traffic-aware adaptive scheduling framework, \texttt{TraDE}, which can adaptively redeploy microservice instances to maintain desired performance amid changing traffic and network conditions within the hosting cluster. We have implemented \texttt{TraDE} as an extension to the Kubernetes platform. Additionally, we deployed realistic microservice applications in a real compute cluster and conducted extensive experiments to assess our framework's performance in various scenarios. The results demonstrate the effectiveness of \texttt{TraDE} in rescheduling running microservices to enhance end-to-end performance while maintaining a high goodput ratio. Compared with the existing method NetMARKS, \texttt{TraDE} outperforms it by reducing the average response time of the application by up to 48.3\%, and improving the throughput by up to 1.2–1.5× across workloads while maintaining a goodput ratio of 95.36\%, and showing robust adaptive capability to meet QoS targets under sustained workloads and dynamic networking conditions.
\end{abstract}

\begin{IEEEkeywords}
Microservice, scheduling, performance optimization
\end{IEEEkeywords}

\section{Introduction}
In the pervasive cloud computing domain, microservices have emerged as a key architecture, revolutionizing how cloud-based applications are designed and implemented. Characterized by the modular and decentralized design, microservices provide good flexibility and scalability, catering to the dynamic demands of modern cloud-deployed applications \cite{ms_survey1, ms_survey2, ms_survey3}. However, this transition also introduces significant challenges in microservice management and performance optimization, particularly microservices with complex workflow and dependencies \cite{sharedMS_AcmTranCS, cost_efficient_MS, ms_placement_iot_TMC, ms_rl_TNSM}. Uncertainty in user requests, combined with varying execution paths across microservices and differing communication delays between nodes, significantly increases the difficulty of ensuring Quality of Service (QoS) compliance for microservices deployed on cluster nodes. 
In cloud environments, numerous microservices are co-located, interact with each other, and are often shared among various applications \cite{Erms_ASPLOS23}\cite{ms_rl_SOCC23}. This can lead to worsened performance degradation due to resource contention and cascading delays in the execution paths of requests. These issues directly affect latency and throughput, both critical for meeting QoS targets.

Although existing works have proposed different methods to improve the running performances of containerized microservices, there are still some limitations to these methods. OptTraffic \cite{OptTraffic_ICPP23} optimizes the traffic transmission of containerized microservices across cluster nodes, which fails to consider the cross-node delays under dynamic networking traffic and also introduces complexity by calculating every dependent container replica pair. NetMARKS\cite{NetMARKS_InfoCOM21} determines Kubernetes pod scheduling by the dynamic network metrics collected by Istio Service Mesh \cite{istio}, which have limitations on analyzing the bidirectional metrics between dependent microservices and also may introduce imbalanced load distributions across the cluster nodes. Other existing methods\cite{ChainFormer_ICSOC23, CoNEXT21_colocationMS, ExtendingNetK8s_CCGRID22, ExtendingNetK8s_ICSOC22} also have similar limitations on tackling dynamic workloads, the awareness of cross-node delays, and imbalanced load distributions in the cluster.

To solve the aforementioned challenges, this paper proposes \texttt{TraDE}, a novel framework that utilizes a traffic and network-aware rescheduling approach. \texttt{TraDE} is designed to adaptively redeploy containerized microservices within the cluster by analyzing real-time traffic stress between dependent microservices along with the variations of cross-node delays. By doing so, it seeks to mitigate QoS target violations amid fluctuating user requests and network variations. To implement and evaluate the proposed \texttt{TraDE} framework, this paper seeks to address several crucial challenges in dealing with network dynamics for meeting application services' QoS targets: (1) How to quantitatively map dynamic bidirectional traffic patterns into traffic stress between upstream and downstream microservices, including all the corresponding replicas?, (2) What approach should be employed to build the traffic stress graph under dynamic workloads within a specific time interval?, (3) How to design a controllable network-dynamics manager to thoroughly evaluate the proposed method via efficiently injecting dynamic cross-node delays and accurately measuring the node-delay matrix?, (4) With the constructed traffic-stress graph and the measured cross-node delay matrix, how to determine the service-to-node mapping under minimal exploration time and a balanced load goal?, (5) With the explored service-to-node mapping result, what strategies should be adopted to ensure zero downtime of the running services when migrating related microservice containers?  


The proposed \texttt{TraDE} framework resolves these challenges and could adapt to the changing traffic conditions and redeploy the running microservice containers to server nodes when QoS violation is detected. Specifically, \texttt{TraDE} builds a traffic stress graph for dependent microservices, a lightweight cross-node delay monitor, a low overhead service-node mapper, and a microservice rescheduler with guarantees of service availability when migrating containers. We demonstrate the effectiveness of our approach through extensive evaluation using practical microservice applications. The results of our experiments, conducted under a variety of scenarios, demonstrate the ability of \texttt{TraDE} to maintain the desired QoS target of deployed microservice applications when QoS violations happen. In summary, our main contributions can be summarized as follows.

\begin{itemize}
    \item We propose a traffic analyzer that dynamically constructs a traffic stress graph. This graph not only illustrates the latest microservice call graphs (dependencies) but also identifies the microservice pairs experiencing higher stress via bidirectional traffic analysis. 


    \item We design and implement a network-dynamics manager which mainly consists of injecting customized cross-node delays in a controllable way via packet-level tagging and also accurately measuring the node communication delays via cluster-level daemon agents.

    \item We introduce a parallel algorithm for service-node mapping to minimize the total traffic transmission overhead with guaranteed balanced task chunks and fast convergence.

    \item We design a microservice rescheduler to migrate microservice instances that experience QoS violations. Specifically, when migrating the microservices, we employ various scheduling schemes to guarantee service availability and resource availability in the cluster. 

    \item We develop a prototype system of the \texttt{TraDE} framework as an extension to the Kubernetes platform, and demonstrate system performance in a real computing cluster.
    
\end{itemize}

\textit{What is new compared to prior work?} TraDE (i) models \emph{bidirectional} traffic to expose overloaded pairs as rescheduling targets; (ii) introduces a \emph{controllable}, destination-specific delay generator paired with a multi-agent measurer for accurate cross-node latency under dynamics; and (iii) couples these with a service-node mapper that embeds an overload penalty and preserves availability during service migrations. Together, these enable fast, practical, and repeatable QoS-target compliance under changing traffic and network conditions.

The rest of the paper is structured as follows: Section 2 discusses related work, providing a comprehensive background on existing methods. The motivation and problem statement are introduced in Section 3. Section 4 presents the proposed \texttt{TraDE} framework, explaining its main components. Section 5 focuses on the design of the traffic analyzer. Section 6 details the design of the dynamics manager, which consists of the dynamic delay generator and the cross-node delay measurer. Section 7 introduces the proposed PGA algorithm for microservice placement, along with an overhead analysis. Section 8 offers a performance evaluation and analysis of the proposed \texttt{TraDE} framework. Finally, Section 9 concludes the paper and discusses future work.



\section{Related Work}
The scheduling of microservices in cluster environments has been extensively studied from various perspectives. This section reviews the existing literature, focusing on microservice management, graph analysis, and network-aware scheduling.

\subsection{Microservice Management}
FIRM \cite{FIRM_MS} leverages online telemetry metrics data and machine learning-based models to manage microservices in a fine-grained way by localizing the SLO violations, identifying resource contentions, and then taking reprovision measures to mitigate the SLO violations. Erms \cite{Erms_ASPLOS23} builds resource scaling models to calculate the latency objectives for shared microservices with large calling graphs. GrandSLAm \cite{grandslam_MS} estimates the completion time of the requests for individual microservice execution stages and leverages the estimated time to batch and reorder the requests dynamically. However, these existing methods have limitations in considering the dynamic impacts of cross-node communication delays and the changing bidirectional traffic between upstream and downstream microservices with multiple replicas.

\subsection{Graph Analysis}
Sage \cite{Sage_ASPLOS21} builds a graphical-based bayesian model to analyze the root cause of cascading QoS violations for interactive microservices focusing on practicality and scalability. Tian et al \cite{TaskDependenceis_SoCC19} develop a workload generator to synthesize the DAG jobs with graph workflow by analyzing large-scale cluster traces. Furtherly, Luo et al \cite{SoCC2021_MS_luo} characterize the call graph of dependent microservices by analyzing Alibaba cluster data and reveal three types of calling dependency graphs for microservice applications. Parslo \cite{Parslo_SoCC21} introduces a gradient descent-based method by breaking the end-to-end SLO budget into smaller units to assign partial SLOs among nodes in a microservice graph under an end-to-end latency SLO. However, these methods are generally time-consuming with high overheads to build the graph and not suitable for dynamic incoming user requests.

\subsection{Network-aware Scheduling}
NetMARKS \cite{NetMARKS_InfoCOM21}  introduces a network-aware approach to schedule the Kubernetes pods from different 5G edge applications by using the collected dynamic metrics from Istio Service Mesh. Marchese et al \cite{ExtendingNetK8s_ICSOC22} introduce a network-aware scheduling extension for the default Kubernetes scheduler by considering the infrastructure network conditions and the interactions among microservices. OptTraffic \cite{OptTraffic_ICPP23} develops a network-aware scheduling framework by optimizing the cross-machine traffic scheduling for multi-replica microservice containers by migrating the containers with dependent relations. However, these networking-aware scheduling schemes designed for microservices still have limitations on bidirectional traffic analysis for dependent microservice replicas and the changing infrastructure-level conditions, i.e. varied cross-node communication delays. 

\section{Motivation and Problem Statement}

\subsection{Background}
An increasing number of modern cloud applications have evolved into microservice-based architectures, which manage applications through a collection of containerized, loosely coupled, fine-grained services \cite{SoCC2021_MS_luo} \cite{deathStarBench_ASPLOS19}. As discussed in previous work \cite{deathStarBench_ASPLOS19}, transitioning from monolithic designs to microservice designs for cloud applications leads to a higher proportion of processing time being spent in the networking stack compared to monolithic applications. The processing time percentages of different cloud applications are shown in Fig. \ref{fig:App_processingTime}. 

\begin{figure}[th]
    \centering
    \includegraphics[width=\linewidth]{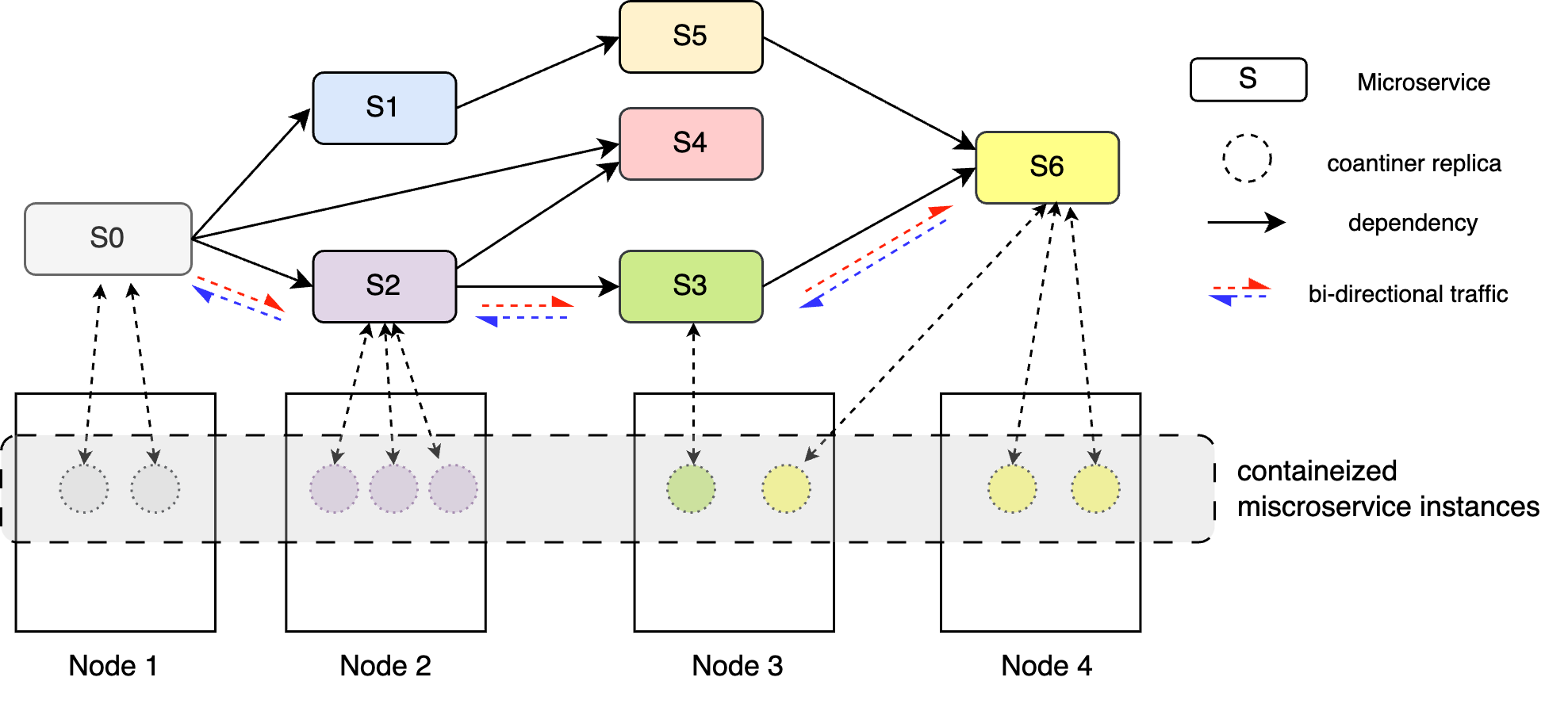}
    \caption{An envisioned workflow of containerized microservice executions and communications in a cluster with four nodes.}
    \label{fig:ms_overview}
\end{figure}

Deployed as a set of containerized microservice instances, the application processes user requests through dependent microservice replicas and returns the processed results in the reverse direction. As shown in Figure ~\ref{fig:ms_overview}, a microservice-based application is decoupled into a collection of containerized microservices along with the bidirectional traffic flows between dependent microservices. Every microservice is supported by one or multiple corresponding container instances to provide the application functionality. These container instances launch from pre-defined container images specified in the corresponding microservice deployment file.


\subsection{Motivation}

\begin{figure}[ht]
    \centering
    \subfloat[QPS=3000.]{
        \includegraphics[width=0.47\linewidth]{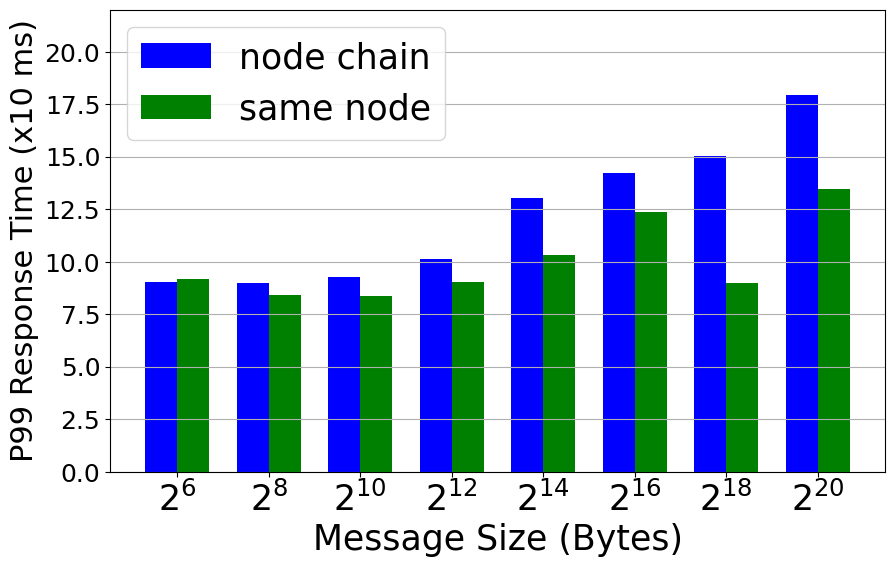}
        \label{fig:QPS_3k}
    }%
    \hfill
    \subfloat[QPS=5000.]{
        \includegraphics[width=0.47\linewidth]{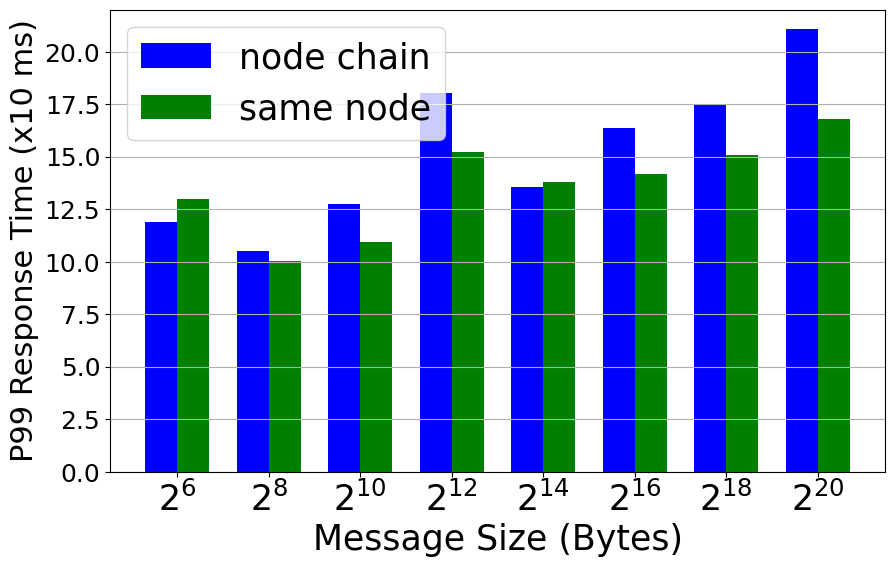}
        \label{fig:QPS_5k}
    }
    \caption{P99 response time of different QPS from an \textit{upstream} microservice client and a \textit{downstream} microservice server under scenarios where they are either colocated on the same node or running across a node chain.}
    \label{fig:crossNodes_diffQPS}
\end{figure}

\begin{figure}[ht]
    \centering
    \subfloat[Processing comparisons of monolithic and microservice applications (data source \cite{deathStarBench_ASPLOS19}).]{
        \includegraphics[width=0.47\linewidth]{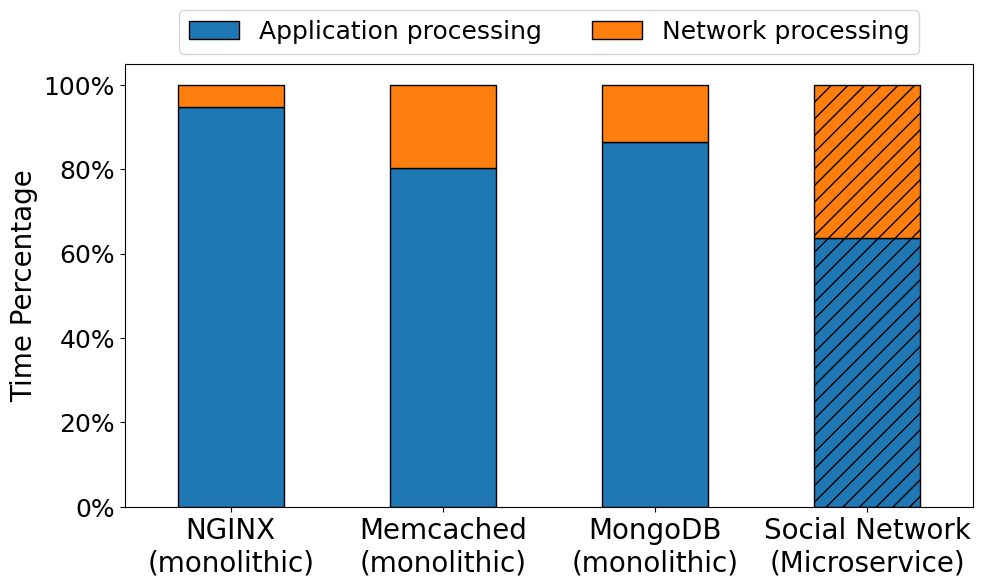}
        \label{fig:App_processingTime}
    }%
    \hfill
    \subfloat[Bidirectional traffic between UM-DM pairs in Social Network in one minute.]{
        \includegraphics[width=0.47\linewidth]{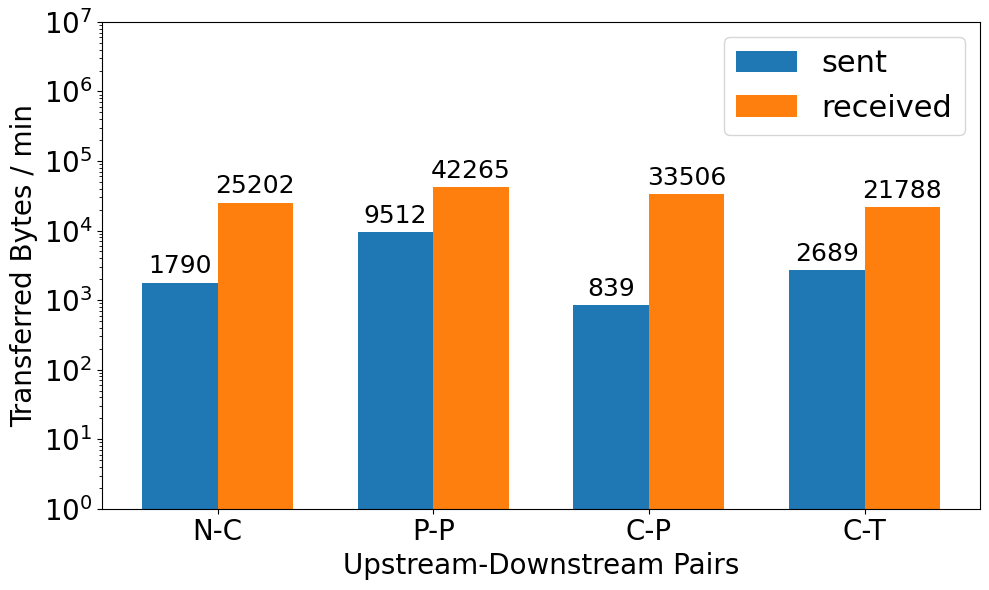}
        \label{fig:UM_DM_tarffic}
    }
    \caption{(a) Comparison of processing time percentage between monolithic and microservice applications. (b) Traffic of different UM-DM pairs in one minute.}
    \label{fig:motivation}
\end{figure}

We deployed synthetic microservice applications in a running cluster to motivate our design and tested them under different scenarios. The deployed applications include a server/client application and a benchmark application. Figure ~\ref{fig:crossNodes_diffQPS} shows the performance of two dependent microservices: one functions as a client container sending \texttt{PUT} requests with varying message sizes from $2^{6}$ Bytes to $2^{20}$ Bytes, and the other operates as the corresponding server container to receive the requests. From Figure ~\ref{fig:QPS_3k} and Figure ~\ref{fig:QPS_5k}, we observe that colocating two dependent microservice containers on the same node notably improves the p99 response time. Additionally, the QPS (Queries Per Second) has a significant impact on dependent microservice containers. As shown in Figure ~\ref{fig:crossNodes_diffQPS}, when QPS increases from 3000 to 5000, the p99 response times for each transferred message size also increase, with a maximum 78.2\% performance degradation when the sent message size is $2^{12}$ Bytes.
This implies that, for an \textit{upstream}-\textit{downstream} pair, the transferred message size, QPS, and cross-node communications have notable impacts on end-to-end performance. 


Additionally, we deployed the Social Network benchmark released with DeathStarBench \cite{deathStarBench_ASPLOS19} to quantify the transmitted traffic difference between dependent UM-DM pairs. Under different request workloads, we observed notable traffic differences between different UM-DM pairs. 

As shown in Figure ~\ref{fig:UM_DM_tarffic}, there are significant differences between sent and received traffic for certain UM-DM pairs, including N-C (\texttt{Nginx}$\rightarrow$ \texttt{Compose-post}), P-P (\texttt{Post-storage-service} $\rightarrow$ \texttt{Post-storage-MongoDB}), C-P (\texttt{Compose-post} $\rightarrow$ \texttt{Post-storage-service}) and C-T (\texttt{Compose-post} $\rightarrow$ \texttt{Text-service}) \cite{deathStarBench_ASPLOS19}. In our testing scenario, for example, the bidirectional traffic between the C-P pair shows that the received traffic could be approximately 40x more than the sent traffic, implying that a slight delay increase in transmission among the C-P (pair could degrade the pair's end-to-end performance by up to 40x. Thus, it is significant to analyze the real-time bidirectional traffic for dependent microservices.

From the above observations, we can conclude that: (1) Unlike monolithic cloud applications, microservice-based applications spend significantly more time on the networking processing stack. (2) Cross-node communication can impact the performance of microservice pairs, particularly for pairs with high volumes of traffic transmission. (3) For dependent \textit{upstream} and \textit{downstream} microservice pairs, the transmitted bidirectional (sent and received) traffic can exhibit significant differences, such as small requests but large payloads.

\subsection{Problem Definition}

In dynamic networking and traffic environments, the deployment of microservice instances across different server nodes plays a critical role in determining the end-to-end performance of distributed applications. In particular, poor placement decisions may incur high communication overhead due to inter-node delays and traffic volumes. To address this, we formulate the optimization task as a \textbf{Service-Node Mapping Problem}, where the objective is to assign a set of microservices to a set of server nodes in a manner that minimizes overall communication cost—capturing both communication latency and traffic-induced overhead—while satisfying server resource constraints.

Let $M=\{m_1,\ldots,m_k\}$ be the set of microservices, $N=\{n_1,\ldots,n_p\}$ the set of server nodes, and $P:M\!\to\!N$ the service–node mapping. Over a measurement window $\Delta t$, let $T_{i\to j}\!\ge\!0$ and $T_{j\to i}\!\ge\!0$ denote the forward and reverse traffic (bytes or rate) between $m_i$ and $m_j$. Let $D_{a,b}\!\ge\!0$ be the one-way (ping-like) delay from node $a$ to node $b$ (not necessarily symmetric). We use direction weights $w_f,w_b\!\in\![0,1]$ (default $w_f{=}w_b{=}0.5$).

The pairwise communication cost for $(i,j)$ under a placement $P$ is
\begin{equation}
\left\{
\begin{aligned}
\mathcal{C}_{i,j}(P) \;=&\; w_f\,T_{i\to j}\,D_{\,P(m_i),\,P(m_j)}
\;+\; w_b\,T_{j\to i}\,D_{\,P(m_j),\,P(m_i)},\\
&(w_f,w_b)\in[0,1]^2.
\end{aligned}
\right.
\label{eq:pair_cost}
\end{equation}
and the total communication cost aggregates pairwise costs:
\begin{equation}
\mathrm{TotalCost}(P)=\sum_{i=1}^{k}\sum_{j=1}^{k}\mathcal{C}_{i,j}(P).
\label{eq:probelm_definition}
\end{equation}

Each node $n\!\in\!N$ has a capacity vector $C_n$ (e.g., CPU, memory, GPU), and each microservice $i$ has a resource demand vector $R_i$. A placement is feasible if, element-wise,
\begin{equation}
\sum_{\,i:\,P(m_i)=n} R_i \;\le\; C_n
\qquad \forall\,n\in N.
\label{eq:resource_constraint}
\end{equation}

\textit{Penalty form used in our solver.}
We enforce \eqref{eq:resource_constraint} via a soft penalty, consistent with our implementation. Let
$\Phi(P)=\sum_{n\in N}\!\max\!\{0,\sum_{i:P(i)=n}R_i - C_n\}$ denote the aggregated overflow (applied element-wise and summed across resources), and let $\lambda>0$ be a large coefficient. The penalised objective is
\begin{equation}
\min_{P}\;\; \mathrm{TotalCost}(P)\;+\;\lambda\,\Phi(P),
\label{eq:penalized}
\end{equation}
which is equivalent in practice to the constrained form for sufficiently large $\lambda$.

\textit{Why bidirectional traffic?}
An RPC is typically a request–response round trip; even with one-way (ping-like) delays, both directions (forward $i{\to}j$, return $j{\to}i$) contribute to user-perceived time and to total network time over $\Delta t$. Equation~\eqref{eq:pair_cost} therefore weights both directions. When one-way delays are symmetric ($D_{a,b}{=}D_{b,a}$), \eqref{eq:pair_cost} reduces to $(w_fT_{i\to j}{+}w_bT_{j\to i})\,D_{\,P(m_i),\,P(m_j)}$; setting $(w_f,w_b){=}(1,0)$ yields the one-direction variant.

\textit{Implementation in TraDE.}
In our prototype system, $T_{i\to j}$ and $T_{j\to i}$ are computed from Istio byte counters over $\Delta t$; we take $w_f{=}w_b{=}0.5$ by default and optimise \eqref{eq:penalized} using a parallel greedy search.

The problem is thus to find a mapping $P$ that minimizes \eqref{eq:probelm_definition} (or equivalently \eqref{eq:penalized}) subject to \eqref{eq:resource_constraint}, promoting traffic locality and reducing cross-node latency under dynamic workloads and network conditions.

\section{Framework of \texttt{TraDE}}
Based on our observations, we were motivated to design \texttt{TraDE}, a network and traffic-aware rescheduling framework for containerized microservices when the deployed service experiences QoS violations due to dynamic requests and varying cross-node communication delays in dynamic computing environments. As shown in Fig. \ref{fig:TraDE}, the figure illustrates the main modules and how each module works together at different stages to complete the adaptive scheduling process. The key modules of the proposed framework are as follows:

\begin{figure}[htbp]
    \centering
    \includegraphics[width=0.98\linewidth]{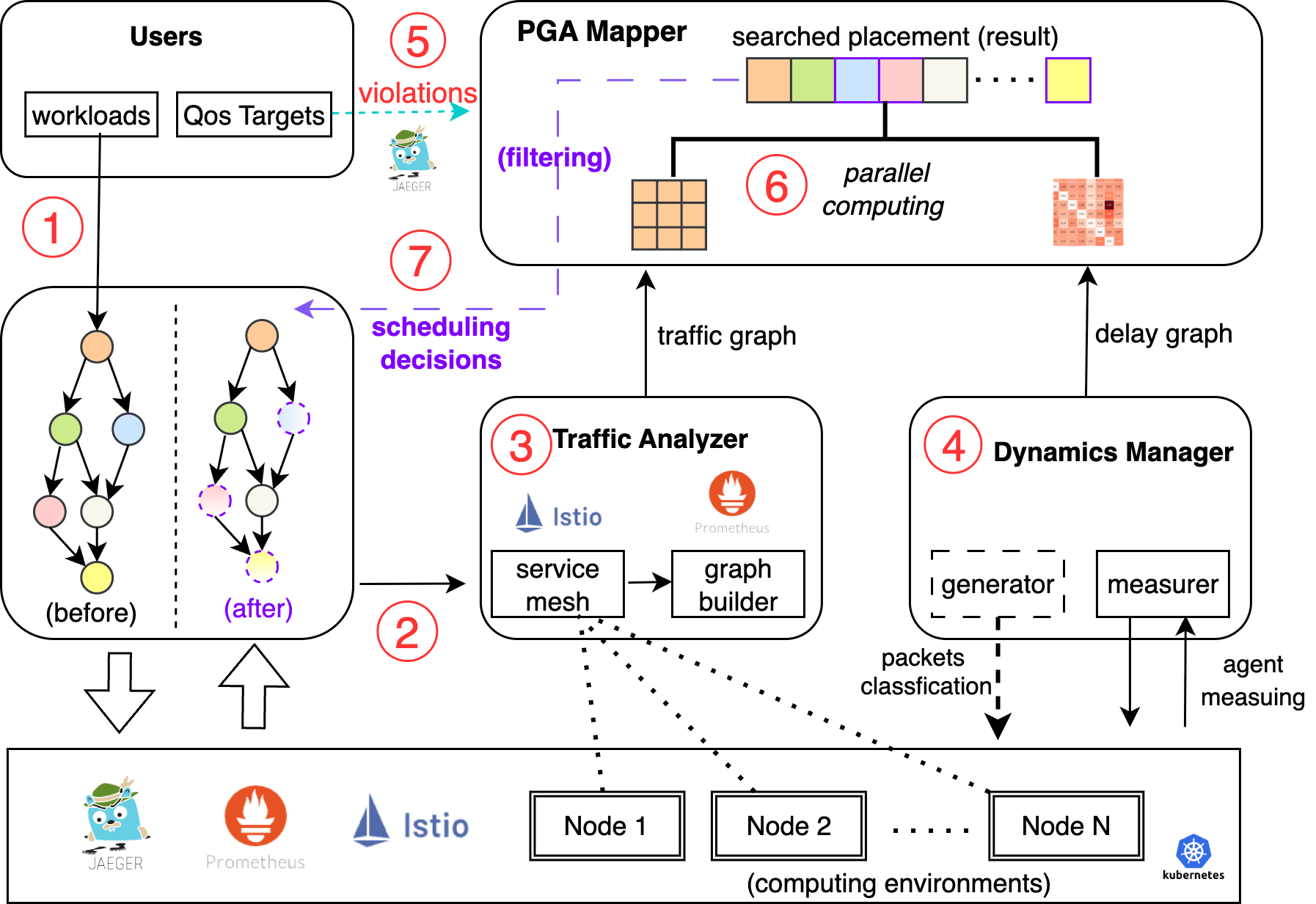}
    \caption{The proposed \texttt{TraDE} framework.}
    \label{fig:TraDE}
\end{figure}
    \begin{itemize}

    \item \textbf{Traffic Analyzer}: This module consists of a service mesh and a graph builder, which are designed for analyzing the real-time bidirectional traffic flows between the dependent \textit{upstream} microservice containers (including all corresponding replicas) and \textit{downstream} microservice containers through the service mesh and the proposed graph builder algorithm.

    \item \textbf{Dynamics Manager}: This module consists of a \textit{generator} and a \textit{measurer} to manage the cross-node communication delays in the computing environments. The \textit{generator} is designed to generate different practical delays to the computing nodes to validate the proposed \texttt{TraDE} through packet-level manipulation. The \textit{measurer} is designed to measure the communication delays between different nodes across the computing environments through multiple daemon agents.
    
    \item \textbf{PGA Mapper}: This module is designed to tackle QoS violations of the running applications. Specifically, when there are violations to the predefined QoS targets, the PGA (Parallel Greedy Algorithm) mapper computes the service placements on the computing nodes to achieve the lowest overhead as defined in Eq.\ref{eq:probelm_definition}. 
    
\end{itemize}

As shown in Fig. \ref{fig:TraDE}, at the beginning stage \textcircled{\raisebox{-0.9pt}{1}}, users define the QoS targets and send different workloads (i.e., different types of requests and QPS) to the deployed microservice applications. At stages \textcircled{\raisebox{-0.9pt}{2}} and \textcircled{\raisebox{-0.9pt}{3}}, the performance metrics of the running microservices are collected by \texttt{Jaeger} and \texttt{Istio}. Within a predefined monitoring time interval, the traffic analyzer analyzes the bidirectional traffic between dependent microservices and builds the traffic graph, which is then sent to the PGA mapper. Meanwhile, at stage \textcircled{\raisebox{-0.9pt}{4}}, the node dynamics manager measures the cross-node communication delay graph, which is sent as another graph to the PGA mapper. As the microservice application runs, if there are any QoS violations (at stage \textcircled{\raisebox{-0.9pt}{5}}) to the predefined QoS targets, \texttt{TraDE} is triggered to run the PGA mapper. As shown at stage \textcircled{6}, the PGA mapper finds the service-node placement through a constructed traffic graph and delay graph from the traffic analyzer and dynamics manager, respectively. Designed to run in a parallel computing manner to speed up the process, the PGA mapper provides the searched placement result, specifying the placement of each microservice to a node. Placement results are filtered before being used as the scheduling decision, as some microservices are already placed in the optimal node, and some microservice instances, such as Jaeger agent instances, are not supposed to be migrated. At the final stage \textcircled{\raisebox{-0.9pt}{7}}, the filtered placement results are taken as the adaptive scheduling decision in response to the current dynamic computing environments.

\section{Design of Traffic Analyzer}
\subsection{Service Mesh}

The key aspect of implementing the Traffic Analyzer for \texttt{TraDE} lies in efficiently and minimally analyzing the bidirectional traffic between dependent container pairs, i.e., the upstream and downstream microservice containers. When there is only one microservice instance for both upstream and downstream services, it is straightforward to collect and analyze bidirectional traffic metrics from the Linux \textit{proc} file system. However, when multiple replicas exist for either the downstream or upstream microservices, analyzing the bidirectional traffic between all upstream and downstream microservice replicas becomes time-consuming. Thus, we implemented the service mesh to better observe and manage the complex traffic flows.

\subsubsection{Istio Service Mesh Implementation} To obtain finer-grained metrics of bidirectional traffic between upstream microservices and their downstream counterparts, along with their corresponding replicas, we implemented \textit{Traffic Analyzer} by analyzing traffic metrics with Istio Service Mesh \cite{istio}. The service mesh is designed as a dedicated infrastructure layer that can be added to containerized microservice applications. With the service mesh deployed, the traffic for each microservice container is proxied by an injected sidecar container.

\begin{figure}[ht]
    \centering
    \subfloat[Service mesh graph.]{
        \includegraphics[width=0.46\linewidth]{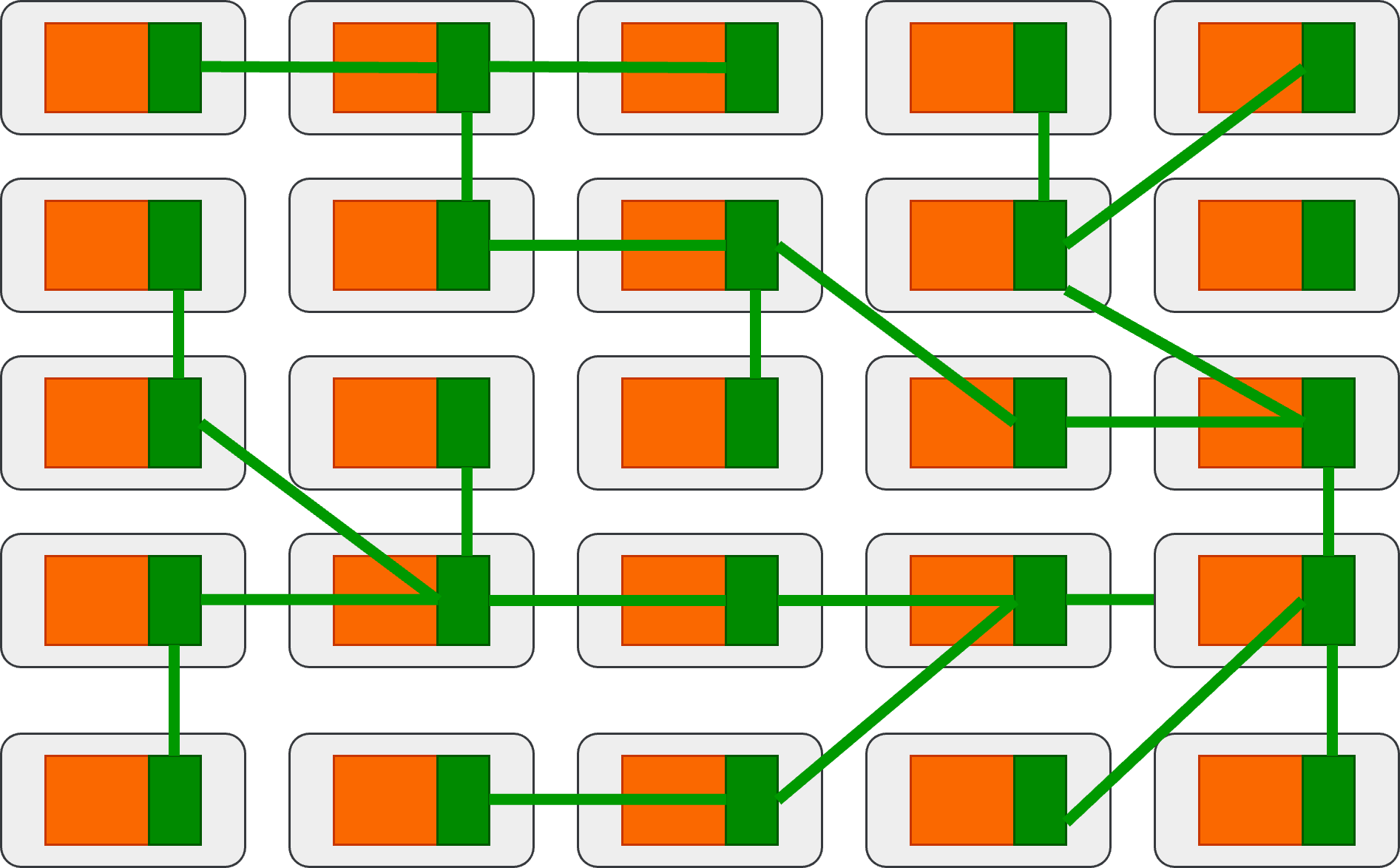}
        \label{fig:serviceMeshGraph}
    }%
    \hfill
    \subfloat[p99 tail latency.]{
        \includegraphics[width=0.47\linewidth]{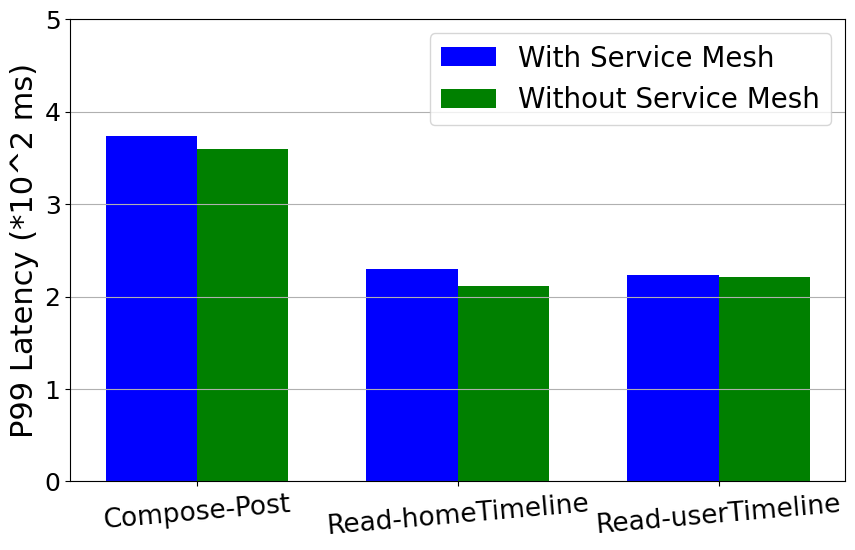}
        \label{fig:MeshPerf}
    }
    \caption{(a) An overview of how service mesh with sidecar containers (green) works with microservice containers (orange). (b) Overhead comparisons of p99 tail latency of different workloads to Social Network applications with and without service mesh.}
    \label{fig:serviceMeshIntro}
\end{figure}

\begin{figure}[t]
    \centering
    \includegraphics[width=\linewidth]{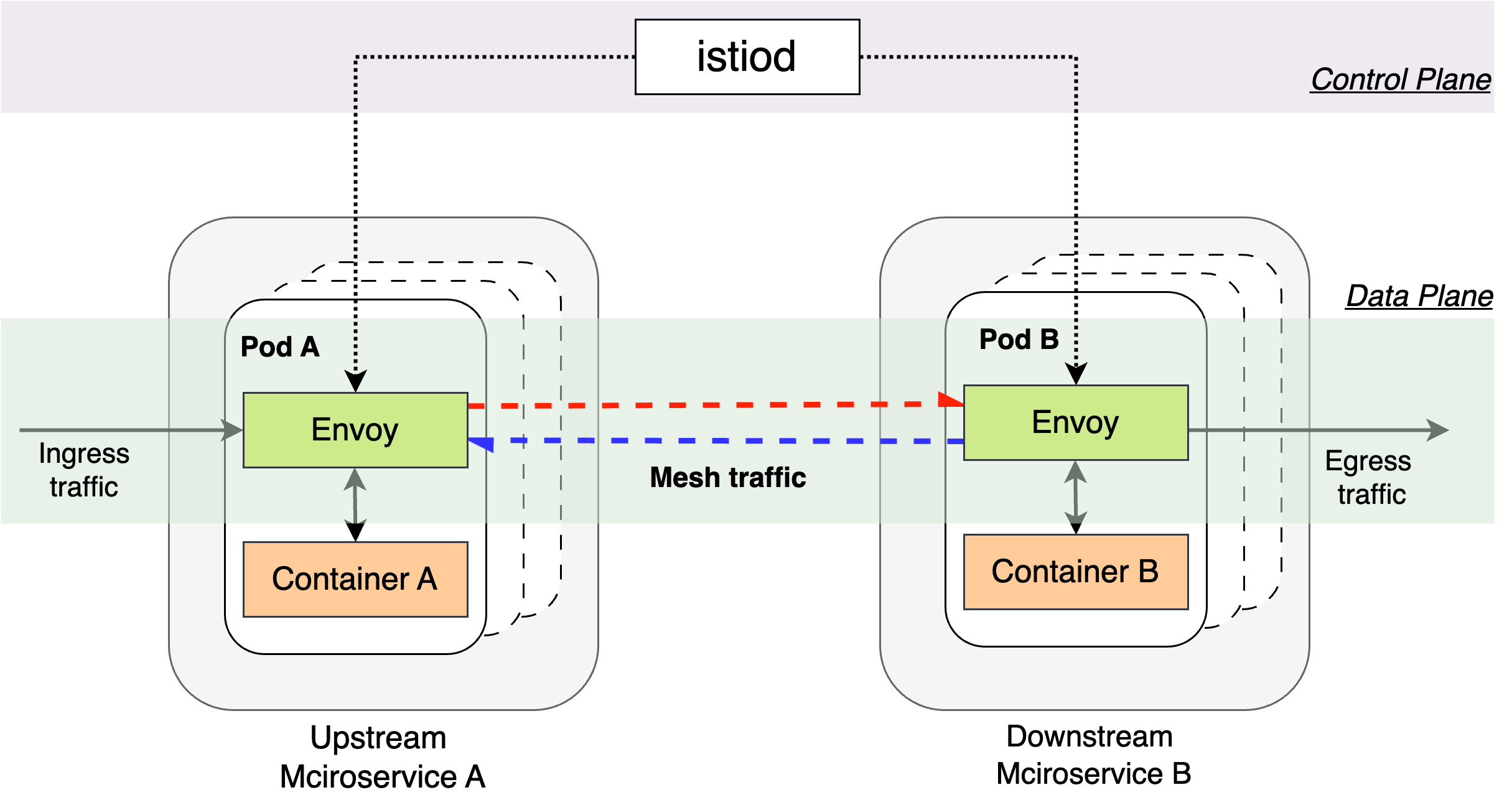}
    \caption{An architecture of how upstream microservice A interacts with downstream microservice B with Istio service mesh enabled.}
    \label{fig:UM_DM_istio}
\end{figure}

In a Kubernetes-based cluster with the deployed Istio service mesh, a microservice pod comprises both the application and sidecar containers. As depicted in Figure ~\ref{fig:serviceMeshGraph}, the dedicated service mesh structure proxies the ingress and egress traffic among the microservice containers. The orange rectangles represent the containerized microservices, and the green rectangles represent the sidecar containers, which function as a dedicated layer managing traffic among various microservices. The green links illustrate the data traffic connections and dependency relationships between microservices. Specifically, as depicted in Figure ~\ref{fig:UM_DM_istio}, the \texttt{Envoy} \cite{Envoy} containers (acting as sidecar containers) in Pod A and Pod B proxy the bidirectional traffic between Container A and Container B. Within each pod, the \texttt{Envoy} container communicates with its corresponding microservice containers.

\subsubsection{Overhead Analysis}
We realize that adding an additional service mesh layer to the microservice application may introduce extra delays in the traffic flow. However, the introduced delay only contributes a slight proportion to tail latency. Based on \cite{istio_perf}, within the Istio 1.21.2 service mesh, utilizing telemetry v2, each request is processed by both a client-side and server-side Envoy proxy. These proxies collectively increase latency at the 90th percentile by approximately 0.182 milliseconds and at the 99th percentile by about 0.248 milliseconds, compared to the baseline latency of the data plane. These findings were conducted with a 1kB payload, a rate of 1000 requests per second, and varying client connections (2, 4, 8, 16, 32, 64) at the CNCF Community Infrastructure Lab \cite{CNCF_lab}.

Additionally, we evaluated and compared the performance of the Social Network \cite{deathStarBench_ASPLOS19} with and without the implemented service mesh. We separately deployed the Social Network application in two different namespaces to guarantee the requests and services would not interfere with each other. Three types of requests (i.e., Compose-Post, Read-homeTimeline, and Read-userTimeline) were sent as the workloads to the deployed Social Network application, each with a QPS of 200. As depicted in Figure~\ref{fig:MeshPerf}, it can be observed that the service mesh introduces only minimal delays. Therefore, the service mesh layer does not add significant overhead to the network traffic of the microservice application. As the tradeoff to better manage the complex traffic flows, the introduced minimal delays are acceptable.

\subsection{Traffic Stress Graph}
One of the main goals of the proposed \texttt{TraDE} is to construct real-time traffic flows along the triggered call graphs of the deployed microservices. Multiple request types and varied QPS would trigger different structures of call-graph with different amounts of traffic flows, which complicates the adaptive scheduling of microservices in response to QoS violations.
In our work, we define the following terminology.

\textit{Stress Element}: Stress Element (SE) defines the dependencies between two dependent microservices, i.e., the UM (Upstream Microservice) and DM (the Downstream Microservice), and the traffic stress between the two corresponding microservices. The traffic stress of a SE is calculated by the average traffic of \textit{sent} and \textit{received} in a given time interval, as we believe that the traffic of \textit{sent} and \textit{received} between the dependent microservices both contribute significant stress to the end-to-end performance of the deployed application. Written mathematically, the stress of a Stress Element can be expressed by:

\begin{equation}
\text{stress}^{\mu}_{\sigma}(\mu^{UM}, \sigma^{DM}, \Delta t) = \frac{\text{Bi-direction\_traffic} (\mu^{UM}, \sigma^{DM})}{2\Delta t}
\label{eq:stree_CE}
\end{equation}

In Eq. \ref{eq:stree_CE}, $\mu$ and $\sigma$ refer to the dependent upstream and downstream microservices. $\Delta t$ is to define the measurement time interval for microservices $\mu$ and $\sigma$, and $\text{Bi-Direction\_traffic}(\mu^{UM}, \sigma^{DM})$ measures the bidirectional traffic transmitted between $\mu$ and $\sigma$ during time interval $\Delta t$. Thus, a Stress Element can be referred to as $SE(\mu^{UM}, \sigma^{DM}, \text{stress}^{\mu}_{\sigma})$. In our implementation, \texttt{TraDE} analyzes the metrics data retrieved from \texttt{istio$\_$tcp$\_$sent$\_$bytes$\_$total} and \texttt{istio$\_$tcp$\_$received$\_$bytes$\_$total} \footnote{Different version of Istio service mesh may use different standard metrics.}, which respectively measures the size of total bytes sent during response and the size of total bytes received during request.



Before analyzing the traffic flow patterns among microservices, it is essential to build a stress graph that represents the dependencies between microservices and the associated stress on those dependencies. Additionally, sorting the microservice pairs from the constructed traffic graph is important for efficient service-node mapping in subsequent sections. Therefore, we propose algorithms for constructing a traffic stress graph and sorting the microservice pairs based on the constructed stress graph.





\subsubsection{Building the Traffic Stress Graph}
Algorithm \ref{alg: stress_graph} constructs the traffic stress graph, \textit{Graph}, using the stress elements defined by Eq. \ref{eq:stree_CE}. The algorithm begins by retrieving a list of deployed microservices at the current monitoring time. It then initializes the \textit{Graph} with zeros for each row and column. In the next step, it iterates through all deployed microservices and calculates the traffic stress between each pair of microservices over a given time interval. Once the matrix iteration is complete, the traffic stress graph is obtained.

\begin{algorithm}[t]
\caption{Build Traffic Stress Graph} \label{alg: stress_graph}
\begin{algorithmic}[1]
\State \textbf{Input:} List of $Stress\_Elements$ with $SE$ objects
\State \textbf{Output:} A traffic $Graph$ with dependencies and stress

\State $MS \gets Stress\_Elements$ \Comment{list of deployed microservices}
\State Initialize $Graph \gets zeros(|MS| \times |MS|)$ 

\For{each $\mu$ in $MS$}            \Comment{upstream microservices}
    \For{each $\sigma$ in $MS$}     \Comment{downstream microservices}
        \If{$\mu \neq \sigma$}
            \State $stress \gets$ \Call{Bi\_traffic}{$\mu$, $\sigma$, $\Delta t$}
            \State $Graph[\mu][\sigma] \gets stress$
        \EndIf
    \EndFor
\EndFor

\State \textbf{return} \textit{Graph}
\end{algorithmic}
\end{algorithm}

\subsubsection{Sorting Microservice Pairs by Stress Level}
After constructing the traffic stress graph, the next step is to identify the stress level among the microservice pairs in the stress graph. Algorithm \ref{alg: sort_pairs} is responsible for sorting all the microservice pairs with traffic values from the stress graph \textit{Graph} in descending order to identify the microservice pairs with the higher stress and also the pairs with lower stress, which will be used for designing scheduling policies in the proposed \texttt{TraDE}. The microservice pairs with higher stress are the pairs that contribute more to the total communication cost in Eq. \ref{eq:probelm_definition}. A quicker localization of the microservice pairs under higher stress would be good for a quicker convergence to find the satisfied microservices ($M$) to nodes ($N$) mapping ($P : M \rightarrow N$).

\begin{algorithm}
\caption{Sort Microservice Pairs by Traffic Stress} \label{alg: sort_pairs}
\begin{algorithmic}[1]
\State \textbf{Input:} Traffic Stress Graph $Graph$ (traffic stress matrix)
\State \textbf{Output:} Sorted list of microservice pairs by traffic stress

\State Initialize $pairs \gets$ \{\} \Comment{empty list of pairs}
\For{each $\mu$ in $Graph$} 
    \For{each $\sigma$ in $Graph[\mu]$} \Comment{corresponding DM}
        \If{$Graph[\mu][\sigma] > 0$}
            \State Append $(\mu, \sigma, Graph[\mu][\sigma])$ to $pairs$ 
        \EndIf
    \EndFor
\EndFor

\State \Call{SortPairs}{$pairs$}  \Comment{sorting in descending order}
\State \textbf{return} Sorted $pairs$
\end{algorithmic}
\end{algorithm}

\section{Design of Dynamics Manager}
\subsection{Dynamic Delay Generator}
\textbf{Customized Delay Generation.}
Injecting different cross-node delays to the cluster nodes is crucial for evaluating the effectiveness of \texttt{TraDE}. However, implementing various communication delays from one source node to multiple destination nodes poses significant challenges. To the best of our knowledge, no existing work has proposed a method to inject customized communication delays in a controllable manner. Some related works \cite{FIRM_MS, delay_ms_TMC22, delay_edge_SPE22, Adaptive_ms_IPDPS21, adaptive_ms_cloudEdge_TPDS21} mention delay injection to server nodes, but these works either use a static delay matrix or implement uniform communication delays for every egress traffic packet on the node's network interface. This approach leads to two main limitations in evaluation experiments: (1) Uniform communication delays from one source node to all destination nodes prevent distinguishing differences between node pairs, as delays from one source node to all other nodes share the same settings; (2) All other networking services not involving correlated nodes are degraded because the injected delays affect all egress traffic.

\begin{figure}[th]
    \centering
    \subfloat[Traffic packets classifying disciplines.]{
        \includegraphics[width=0.47\linewidth]{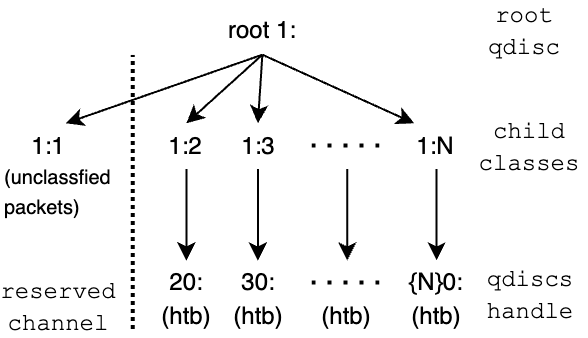}
        \label{fig:qdisc}
    }%
    \hfill
    \subfloat[Packets egress with different class disciplines.]{
        \includegraphics[width=0.47\linewidth]{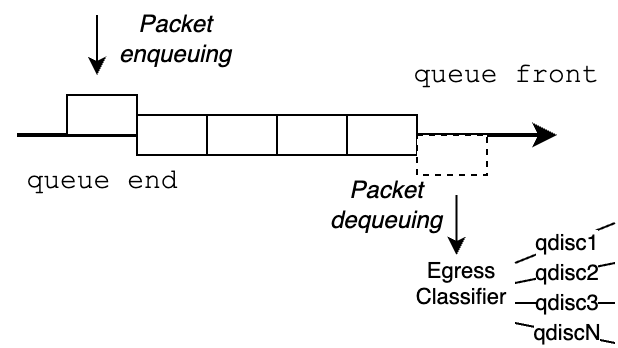}
        \label{fig:packetQueue}
    }
    \caption{(a) Unclassified packets (not sent to certain destinations) are transmitted through a reserved channel without injected delays, while classified packets are assigned different delays for different destinations. (b) Classified packets are sent to different destinations with varying delays.}
    \label{fig:qidsc_classification}
\end{figure}

To address these limitations in current research, as shown in Fig. \ref{fig:qdisc}, we propose a customized delay injection scheme that classifies packets using filters to differentiate egress packets and distribute them based on their IP destinations. Additionally, an extra channel is reserved for default packet transmission without injected delays, ensuring that the performance of other services is not affected. We implemented this scheme using the \texttt{tc} networking tool and \texttt{htb} (Hierarchical Token Bucket).

\subsection{Cross-node Delay Measurer}
In a large-scale computing cluster with high traffic, communication delays among infrastructure nodes are not negligible, as microservice applications are decoupled and distributed across different connected computing nodes. When significant delays occur among cluster nodes, the communication among microservice containers running on these nodes will be negatively impacted. Consequently, some microservices may experience QoS violations.

\begin{figure}[t]
    \centering
    \includegraphics[width=\linewidth]{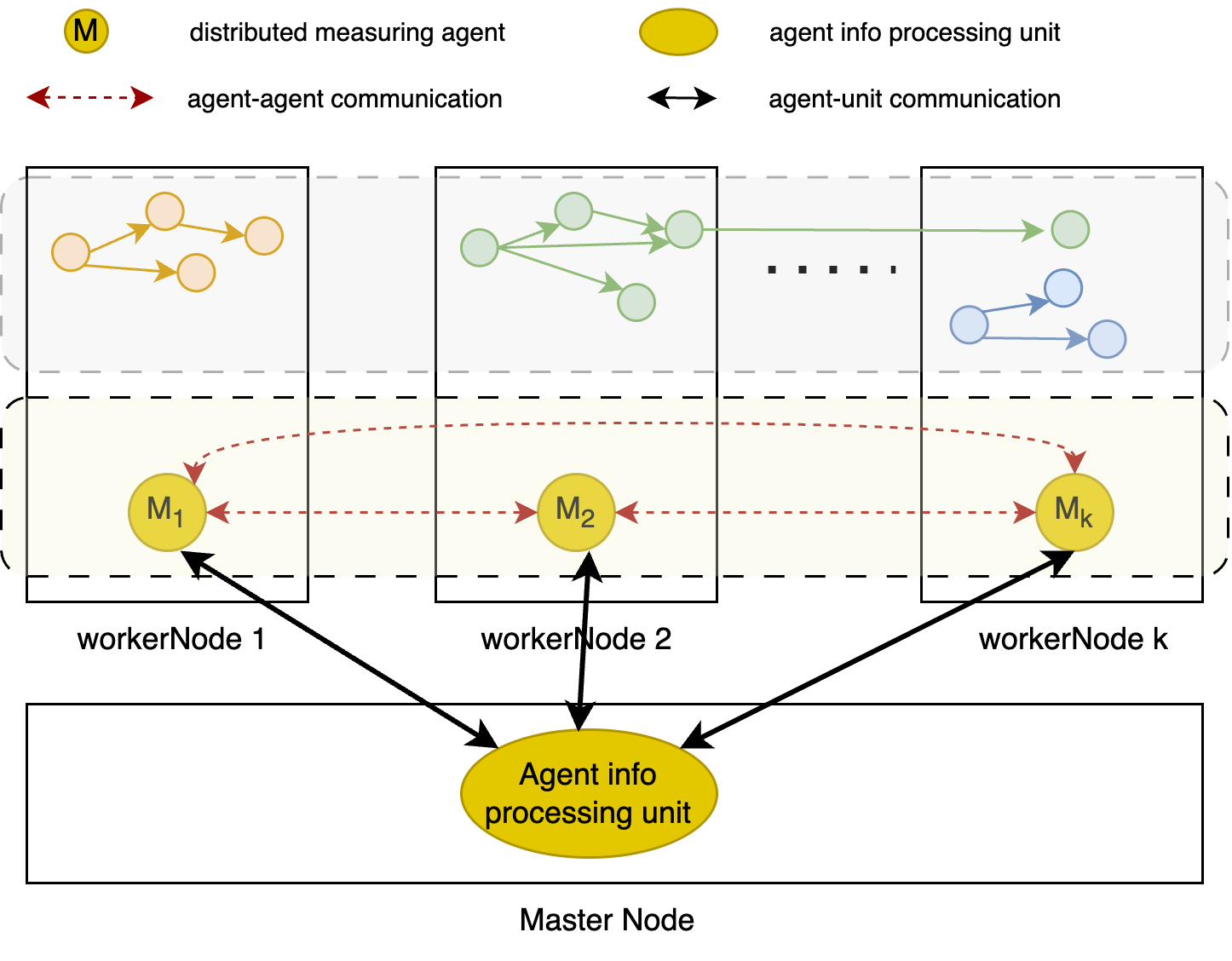}
    \caption{The implementation of \textit{Cross-node Delay Measurer} across cluster nodes.}
    \label{fig:NodeDelay_monitor}
\end{figure}

\textbf{Design of Delay Measurer.} To address this problem, we designed a lightweight module called the \textit{Cross-node Delay Measurer} for our proposed \texttt{TraDE} framework. The module consists of a centralized information processing unit and multiple distributed measuring agents. To keep the module lightweight, the processing unit maintains minimal connections with all measuring agents and summarizes the measured communication delays. The measuring agents run as lightweight containers, maintaining only a simple communication function for measuring delays. Additionally, to enhance the module's robustness during cluster-level upgrades or downgrades, an auto-scaling mechanism for the distributed measuring agents is implemented. Specifically, when the cluster adds more nodes, a measuring agent is automatically added to the new nodes to maintain the consistency and accuracy of communication delay measurements across the cluster. Similarly, when the cluster removes nodes, the corresponding measuring agents are automatically removed as well.




\begin{table*}[!t]
\centering
\caption{Injected delay versus measured cross-node communication delay (ms). In each cell, the first number represents the injected delay, and the second number represents the measured actual delay. This table demonstrates the successful injection of different delays from one source node to various destination nodes, as well as the reserved channel for destinations (i.e., Master node and google.com) without injected delays.}
\label{tab:src_dst_delays}
\resizebox{\textwidth}{!}{%
\begin{tabular}{cccccccccccc}
\toprule[1.2pt]
\textbf{Source Node} & \multicolumn{11}{c}{\textbf{Destinations}} \\ \hline
& \textbf{Node1} & \textbf{Node2} & \textbf{Node3} & \textbf{Node4} & \textbf{Node5} & \textbf{Node6} & \textbf{Node7} & \textbf{Node8} & \textbf{Node9} & \textbf{Master Node} & \textbf{google.com} \\ \hline \hline
\textbf{Node1} & - / - & 3.00 / 3.21 & 8.00 / 8.51 & 10.00 / 11.01 & 14.00 / 14.21 & 6.00 / 6.73 & 27.00 / 28.98 & 13.00 / 13.89 & 21.00 / 22.31 & - / 0.64 & - / 14.10 \\ 
\textbf{Node2} & 8.00 / 8.22 & - / - & 4.00 / 4.52 & 13.00 / 14.02 & 14.00 / 15.21 & 18.00 / 19.02 & 38.00 / 38.22 & 31.00 / 32.13 & 29.00 / 1.02 & - / 0.89 & - / 14.20 \\ 
\textbf{Node3} & 4.00 / 4.03 & 12.00 / 12.37 & - / - & 8.00 / 9.05 & 18.00 / 18.66 & 11.00 / 11.89 & 4.00 / 5.75 & 12.00 / 13.41 & 15.00 / 16.89 & - / 0.43 & - / 14.20 \\ 
\textbf{Node4} & 17.00 / 17.04 & 15.00 / 15.11 & 7.00 / 7.21 & - / - & 5.00 / 5.87 & 6.00 / 6.79 & 22.00 / 23.15 & 13.00 / 13.76 & 25.00 / 26.82 & - / 0.51 & - / 14.30 \\ 
\textbf{Node5} & 20.00 / 21.02 & 12.00 / 12.23 & 11.00 / 11.68 & 10.00 / 11.27 & - / - & 9.00 / 9.73 & 7.00 / 7.13 & 4.00 / 4.11 & 9.00 / 9.23 & - / 0.45 & - / 14.10 \\ 
\textbf{Node6} & 17.00 / 17.12 & 26.00 / 26.06 & 18.00 / 18.87 & 16.00 / 17.08 & 6.00 / 6.11 & - / - & 5.00 / 5.71 & 10.00 / 10.19 & 5.00 / 5.86 & - / 0.53 & - / 14.20 \\ 
\textbf{Node7} & 20.00 / 20.96 & 10.00 / 10.52 & 10.00 / 10.91 & 9.00 / 9.86 & 11.00 / 12.92 & 5.00 / 5.67 & - / - & 5.00 / 5.71 & 9.00 / 9.39 & - / 0.45 & - / 14.10 \\ 
\textbf{Node8} & 21.00 / 21.11 & 25.00 / 25.72 & 4.00 / 4.53 & 10.00 / 10.71 & 12.00 / 12.73 & 15.00 / 16.02 & 10.00 / 10.13 & - / - & 6.00 / 7.08 & - / 0.34 & - / 14.10 \\ 
\textbf{Node9} & 36.00 / 36.93 & 22.00 / 22.70 & 40.00 / 40.39 & 9.00 / 10.08 & 25.00 / 25.69 & 8.00 / 8.16 & 7.00 / 7.23 & 6.00 / 6.18 & - / - & - / 0.30 & - / 14.30 \\ 
\bottomrule[1.2pt]
\end{tabular}%
}
\end{table*}

\textbf{Implementation of Delay Measurer.}
We implemented an efficient, lightweight, and auto-scalable cluster-level measuring scheme that analyzes cross-communication delays across infrastructure nodes. As shown in Fig. \ref{fig:NodeDelay_monitor}, we introduced the design schemes of the \textit{Cross-node Delay Measurer}, consisting of an agent information processing unit and a set of distributed measuring agents.

Specifically, we designed the agent information processing unit to operate as a running plugin on the master node, collecting measured cross-node delay information from all active agents via \texttt{TCP} messages. Meanwhile, the distributed measuring agents are implemented as a group of interconnected running pods, managed by a DaemonSet deployment with a lightweight pre-configured image. In this way, the measuring agents on worker nodes continuously measure cross-node delays with each other and send the measured information to the information processing unit plugin on the master node. Additionally, when the number of cluster nodes changes, the agents are automatically added when a new node joins or deleted when a node drains from the existing cluster through the DaemonSet deployment mechanism. With this mechanism, the \textit{Cross-node Delay Measurer} ensures that each node will have a dedicated pod for delay measurement, regardless of changes in the number of cluster nodes.
 
Fig. \ref{fig:NodeDelay_monitor} shows how the distributed measuring agents communicate with each other and how the agent information processing unit interacts with each measuring agent container.

\textbf{Overhead analysis of Delay Measurer.}
Regarding the overhead of delay measurement in a real system, we optimized the measurement using parallel processing. The latency measurement tasks are designed to run concurrently, and the results are aggregated into a latency results dictionary, completing the process in just a few seconds. 
Additionally, the distributed measuring agents are lightweight (about 0.2 MiB) and stable (running over 6 months without any failure). Each node runs a measuring agent developed from the \texttt{curlimages/curl} image, consuming around 0.2 MiB of memory per node.
Thus, the cluster-level delay measurement consumes minimal memory and completes within a
few seconds.


\subsection{Effectiveness of Dynamics Manager}
Injecting different communication delays from one source node to various destination nodes and accurately measuring these delays is challenging. By adopting the proposed injection scheme shown in Fig. \ref{fig:qidsc_classification}, we injected different delays to various destination worker nodes and measured the actual communication delays from the source node to these destinations. As shown in Table \ref{tab:src_dst_delays}, the measured delays in each cell exhibit high accuracy, with only around a 1 $ms$ difference, which can be attributed to the randomness of data packet transfers in cloud networking stacks. Additionally, to demonstrate that other traffic is not influenced, we tested the communication delays from all worker nodes to other destinations. One destination is the master node in the same cluster, and the other is the Google host (google.com), showing average delays of 0.50 $ms$ and 14.20 $ms$, respectively.

From these measured data, we can confirm the effectiveness of the proposed schemes and implementation for the \textit{Dynamics Manager} in \texttt{TraDE}. It should be noted that the delay injection scheme is optional and can be switched off when \texttt{TraDE} is deployed in actual computing environments. The primary design aim of the customized delay injection scheme is to generate various dynamics to validate and evaluate our proposed adaptive scheduling framework \texttt{TraDE}. Additionally, the proposed delay injection mechanism with customized delays can be easily adopted to evaluate systems in different computing environments, such as cloud-edge continuum and pervasive computing.


\section{PGA Mapper for Microservice Placement}

To solve the defined problem in Section 3.3, we proposed the following PGA (Parallel Greedy Algorithm) parallel algorithm to address the problem. We propose a Parallel Greedy Algorithm to optimize microservice placement by leveraging parallel computing to minimize communication costs. This method ensures efficient and effective placement of microservices, reducing inter-node communication overhead and improving system performance.

\begin{figure}[htbp]
\centering
\begin{lstlisting}[style=calltree]
ParallelPlace(T, D, P, res, cap, workers)
|-- Sorted_MS_pairs(T)
|-- for each chunk task:
|   |-- PlaceWorker(T, D, P, res, cap, tasks)
|   |   |-- CalcCost(T, P, D, res, cap)
|   |   |-- CalcCost(T, new_P, D, res, cap)
|-- Choose best result from all workers
\end{lstlisting}
\caption{Function call hierarchy of the PGA placement Algorithm ~\ref{alg: PGA}.}
\label{fig:function-call-tree}
\end{figure}

\begin{algorithm}
\caption{PGA Algorithm for Microservice Placement under Dynamic Traffics and Cross-node Delays.}\label{alg: PGA}
\begin{algorithmic}[1]
\State \textbf{Input:} Matrix \( T \) for microservice traffic stress graph, Matrix \( D \) for cross-node delay graph, Placement \( P \) for service to node mapping list.
\State \textbf{Output:}  \textcolor{black}{Microservice-node} \(placement \), Lowest \( cost \)
\State \textbf{Define} $res\_list \gets$ \{\texttt{cpu}, \texttt{memory}, \texttt{gpu}, \dots\}

\State $res \gets \Call{get\_ms\_demands}{res\_list, T}$
\State $cap \gets \Call{get\_node\_capacities}{res\_list}$

\State \texttt{// Cost and Overloads Penalty.}

\Function{CalcCost}{$T, P, D, res, cap$}
    \State $cost \gets 0$
    \ForAll{$(u,v)$ in $T$}
        \State $cost \gets cost + T[u][v] \times D[P[u]][P[v]]$
    \EndFor
    \State Initialize penalty factor $pf$ \Comment{$pf$ can be adaptive}
    \State $loads \gets [0] \times \text{len}(cap)$
    
    \For{each $u$ in $P$} \Comment{Calculate server resource load}
        \State $loads[P[u]] \gets loads[P[u]] + res[u]$
    \EndFor
    
    \State $penalty \gets 0$
    \For{each server $j$ in $loads$} \Comment{Check for overloads}
        \If{$loads[j] > cap[j]$}
            \State $penalty \gets penalty + (loads[j] - cap[j]) \times pf$
        \EndIf
    \EndFor

    \State \Return $cost + penalty$
\EndFunction
\State \texttt{// Microservice Placement Worker.}
\Function{PlaceWorker}{$T, D, P, res, cap, tasks$}
    \State $current\_cost \gets \Call{CalcCost}{T, P, D, res, cap}$
    \State $nodes\_num \gets |D|$ \Comment{number of server nodes}
    \For{each $(u, v, stress)$ in $tasks$}
        \State $new\_P \gets$ other nodes for $u$ and $v$
        \State $\_cost \gets \Call{CalcCost}{T, new\_P, D, res, cap}$
        \If{$\_cost < current\_cost$}
            \State Update $P$ and $current\_cost$
        \EndIf
    \EndFor
    \State \Return $P, current\_cost$
\EndFunction

\State \texttt{// Parallel Placement Computing.}
\Function{ParallelPlace}{$T, D, P, res, cap, workers$}
    \State $pairs \gets \Call{Sorted\_MS\_pairs}{T}$ \Comment{Algorithm \ref{alg: sort_pairs}}
    \State $num\_pairs \gets |pairs|$ \Comment{Get the number of pairs}
    \State $num\_workers \gets |workers|$ 
    \State $size \gets \lceil \frac{num\_pairs + num\_workers - 1}{num\_workers} \rceil$

    \State \texttt{/* Distribute tasks to workers */}
    {\color{black}
    \State Initialize $tasks \gets [ ]$ \Comment{Initialize chunk tasks}
    }
    \For{$i = 0$ \textbf{to} $num\_workers - 1$}
        \State $tasks \gets pairs[i \times size : (i + 1) \times size]$
    \EndFor

    \State $results \gets$ \Call{PlaceWorker}{$T, D, P, res, cap, tasks$}
    \State Choose the best $\{P, cost\}$ from $results$
    \State \Return $best\_P, best\_cost$ 
\EndFunction

\end{algorithmic}
\end{algorithm}

\subsection{PGA Algorithm Explanation}
The algorithm aims to iteratively refine the placement of microservices to minimize the total communication cost, as defined in Equation ~\ref{eq:probelm_definition}. It leverages parallel processing to handle multiple microservices concurrently, improving the efficiency of the optimization process. The algorithm integrates three core functions: calculating communication costs, optimizing placements for chunks of microservices, and iteratively refining the overall placement until no further improvement is achievable. \textcolor{black}{As shown in Figure~\ref{fig:function-call-tree}, the \textit{ParallelPlace} function coordinates the execution by calling \textit{PlaceWorker} on distributed microservice pairs, which in turn invokes \textit{CalcCost} to evaluate each candidate placement.}

\textcolor{black}{The key steps of the proposed method for microservice placement are as follows:}

\textbf{1) Initialization}: The algorithm begins with an initial placement of microservices across server nodes, alongside the input of a traffic stress graph and a cross-node communication delay graph. 

\textbf{2) Resource List Definition}: As outlined in Eq. \ref{eq:resource_constraint}, resource constraints such as CPU, memory, and GPU availability are critical when migrating microservices. A resource list \texttt{res\_list} is defined at the start, specifying the resources to be considered throughout the placement process.

\textbf{3) Cost Calculation}: The total communication cost for the initial placement is computed using the \textit{CalcCost (T, P, D, res, cap)} function. This function also accounts for server overloads by introducing a penalty factor, which adds to the communication cost if any server's capacity is exceeded.

\textbf{4) Parallel Processing}: The set of microservices is divided into chunks, and the \textit{PlaceWorker} function is applied to each chunk in parallel via \textit{ParallelPlace} function. 
    \begin{itemize}
        \item \textcolor{black}{In the \textit{PlaceWorker} function, for each microservice pair \((u, v)\), the algorithm attempts to reassign \(u\) and \(v\) to every possible pair of server nodes (excluding their current assignments) across the cluster. This is done exhaustively over all node pairs \((i, j)\), ensuring that all candidate placements are evaluated. The new placement is only accepted if it results in a strictly lower total cost (including penalty for resource violations) compared to the current placement.
        }
        \item \textcolor{black}{In the \textit{ParallelPlace} function, the full set of sorted microservice communication pairs is first partitioned into disjoint chunks, where each chunk is assigned to a separate worker (e.g., CPU thread). Each worker executes its own instance of the \textit{PlaceWorker} function independently. These workers operate in parallel, concurrently exploring local placement refinements for their assigned microservice pairs. After all workers (CPU threads) complete, their outputs—each representing a locally optimized placement and cost—are gathered, and the best result (i.e., the one with the lowest overall cost) is selected as the final placement. This design improves scalability by enabling simultaneous local search across multiple parts of the system.}
    \end{itemize}

\textbf{5) Iterative Refinement}: The results from all parallel workers (i.e., CPU cores) are gathered, and the overall placement is updated if a better solution is found. This process continues until no further improvement can be made in the placement.

\textbf{6) Output}: The algorithm concludes by returning the \textcolor{black}{satisfied} microservice placement across the server nodes, along with the minimized communication cost.

This approach leverages parallel processing to efficiently handle large-scale microservice deployments, ensuring efficient placement with reduced communication overhead.

\subsection{Balanced Chunks and Fast Convergence} \label{ref_balacnedLoad}
\textbf{Balanced Task Chunks}: To balance the computation tasks at each parallel worker, we adopted the chunk size
calculated by $size = \lceil \frac{num\_pairs + num\_workers - 1}{num\_workers} \rceil$
to ensure a more accurate and even distribution of tasks among workers, especially when the number of tasks is not perfectly divisible by the number of workers. This approach correctly handles the remainder and avoids overestimating the chunk size, ensuring a more balanced task load distribution for the parallel computing process.

\textbf{Fast Convergence}: 
Not all microservices contribute equally to the total communication cost. Some pairs of microservices may have significantly higher traffic between them compared to others.
As shown in line 40 of the Algorithm \ref{alg: PGA}, high-traffic microservice pairs will be prioritized. By focusing on these high-traffic pairs first, the algorithm addresses the microservice placements that contribute the most to the total cost. Thus, improvements in these placements will have a disproportionately large impact on reducing the overall cost compared to optimizing low-traffic pairs.

\subsection{Overhead Analysis}
\begin{figure}[t]
    \centering
    \subfloat[PGA Overhead Heatmap.]{
        \includegraphics[width=0.46\linewidth]{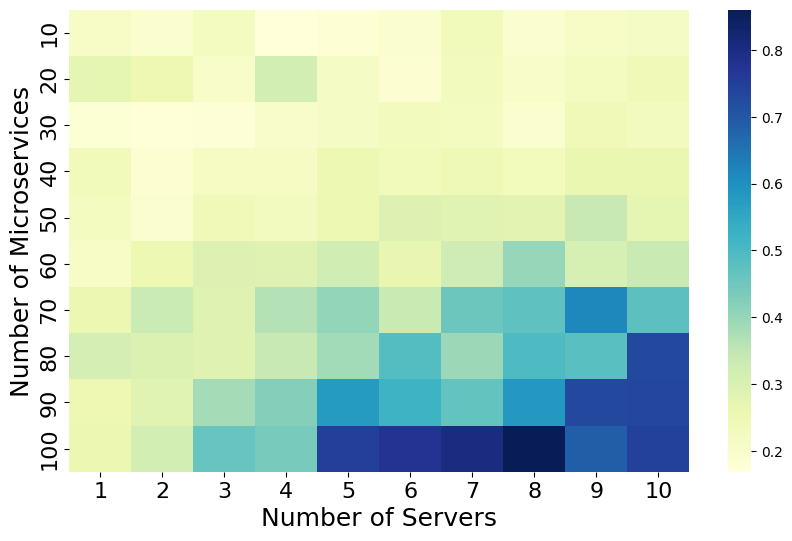}
        \label{fig:pga_overhead_matrix}
    }%
    \hfill
    \subfloat[PGA Overhead Distributions.]{
        \includegraphics[width=0.47\linewidth]{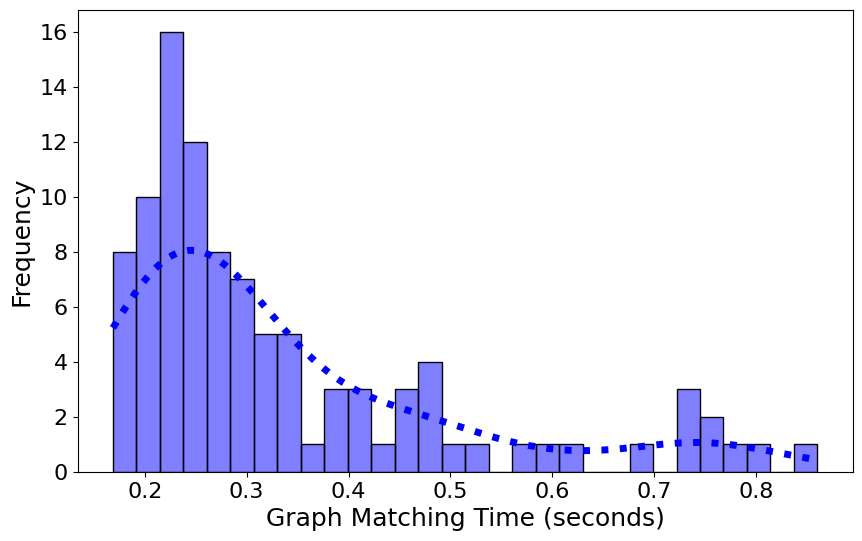}
        \label{fig:pga_overhead_dist}
    }
    \caption{Processing time overheads for exploring \textcolor{black}{satisfied} placements via the PGA algorithm under different numbers of server nodes and microservices.}
    \label{fig:pga_overheads}
\end{figure}

Exploring the \textcolor{black}{satisfied} placement result is time-consuming. To address this, we designed and implemented the proposed algorithm as a parallel algorithm and quantified the overheads in our cluster. When high-impact microservice pairs are grouped and processed in parallel, it reduces the overhead of synchronization between parallel tasks, thus reducing the need for frequent inter-worker communications. As shown in Fig. \ref{fig:pga_overhead_matrix}, the running time of the proposed PGA algorithm is able to complete the matching process in less than 1 second. In Fig. \ref{fig:pga_overhead_dist}, it can be observed that the majority of the graph matching time is approximately 0.3 seconds, which indicates a promising scheduling decision time with a quick response to dynamic changes.

\subsection{Adaptive Scheduler}
\textit{Adaptive Scheduler} is responsible for rescheduling the deployed microservices containers based on the \textcolor{black}{satisfied} placement results from PGA mapper. When there are traffic stresses on the running application or notable communication delays among certain server nodes, the pre-defined QoS targets might be violated in practical computing environments. Thus, designing adaptive scheduling schemes to tackle QoS violations is significant. 

The rescheduling process involves microservice instance migration and eviction across different server nodes in the cluster. However, this will lead to the following problems:  (1) how to guarantee the performance of microservice instances that are not affected during the rescheduling process?, (2) when service consistency is guaranteed, how to avoid the overloads on certain server loads? 

\textbf{Asynchronous launching}: At the beginning of microservice migration, it is crucial to ensure the service availability of the affected microservices. We initially launch new microservice containers on the target cluster node, and once these new containers are in ready states, the old containers are evicted from the previous nodes. This approach guarantees zero downtime for specific microservices, whose backend-supported microservice instances require migration.

\begin{figure}[th]
    \centering
    \includegraphics[width=\linewidth]{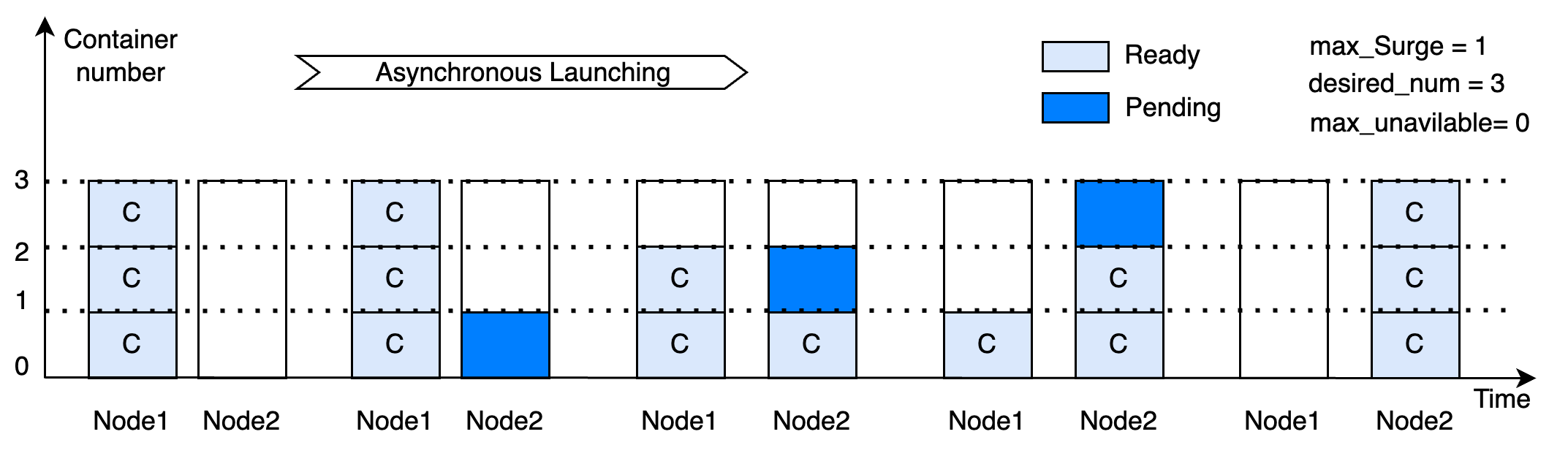}
    \caption{The process of asynchronous launching for container migration (evicting old while launching new containers) between two nodes to guarantee service high availability.}
    \label{fig:container_migration}
\end{figure}

\textbf{Constraints-based scheduling}: \textcolor{black}{The proposed rescheduling scheme computes a service–node placement based on microservice dependencies and the cross-node delay matrix, and also enforces server-node resource constraints (e.g., CPU, memory, GPU), as mathematically summarized in Eq. \ref{eq:resource_constraint}.}
In line 2 of the Algorithm \ref{alg: PGA}, a resource constraints list is defined for consideration during the whole rescheduling process. Additionally, the microservice demands (e.g., \texttt{requests} and \texttt{limits} in deployment yaml file) and server node resource capacities are considered as conditions at Algorithm \ref{alg: PGA}.

\section{Performance Evaluation}
\subsection{Testbed Setup}
\textbf{Cluster Setup}: We validated our design by implementing \texttt{TraDE} using the de facto standard container orchestration platform, Kubernetes \cite{kubernetes}. We deployed \texttt{TraDE} on 10 server nodes without any preset anti-colocation rules, such as taints for server nodes and affinity for pods. The Kubernetes cluster configuration includes one master node and nine worker nodes. The master node features 32 CPU cores with x86\_64 AMD EPYC 7763 series processors, 32GiB of RAM, and a network bandwidth of 16 Gbps. Each of the nine worker nodes is equipped with 4 CPU cores from the same AMD series as the master node, 32 GiB of RAM, and a network bandwidth of 16Gbps. Regarding software versions, the cluster runs Kubernetes v1.27.4, the Container Network Interface (CNI) plugin Calico v3.26.1, the service mesh Istio v1.20.3, and uses CRI-O v1.27.1 as the container runtime. All server nodes operate on Ubuntu 22.04.2 LTS with the Linux kernel 5.15.0.

Besides, each of the cluster nodes is running on a Virtual Machine instance at the dedicated research cloud platform from the University, thus the typical communication delay among each of the nodes is ultra-low, usually ranging from 0.2 to 1 milliseconds ($ms$), tested by sending \texttt{ICMP} messages between nodes. To make our evaluations more realistic and emulate the changing cluster networking environment, the cross-node delay matrix can be automatically updated every $t$ (i.e., $t=5$) minutes. In Table \ref{tab:src_dst_delays}, we demonstrated the customized cross-node communication delays to the worker node destinations and also avoided injecting delays to other destinations like the master node and external sites.

\textbf{Benchmark Application}:
We adopted \texttt{Social Network} from DeathStarBench\cite{deathStarBench_ASPLOS19} as the microservice application to evaluate our proposed scheduling framework. \texttt{Social Network} benchmark emulates a simplified social media platform similar to popular social networking services. It is structured to replicate the intricate interactions and communication patterns in such type of applications. The benchmark comprises 27 different microservices that collectively offer functionalities such as composing a post, writing a user timeline, and writing a home timeline.

\begin{figure*}[t]
    \centering
    \subfloat[\texttt{compose-post}.]{
        \includegraphics[width=0.24\textwidth]{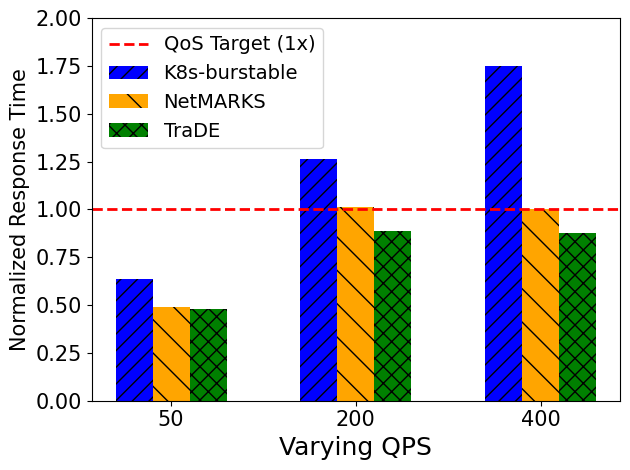}
        \label{fig:rt_compose-post}
    }%
    \hfill
    \subfloat[\texttt{read-user-timeline}.]{
        \includegraphics[width=0.24\textwidth]{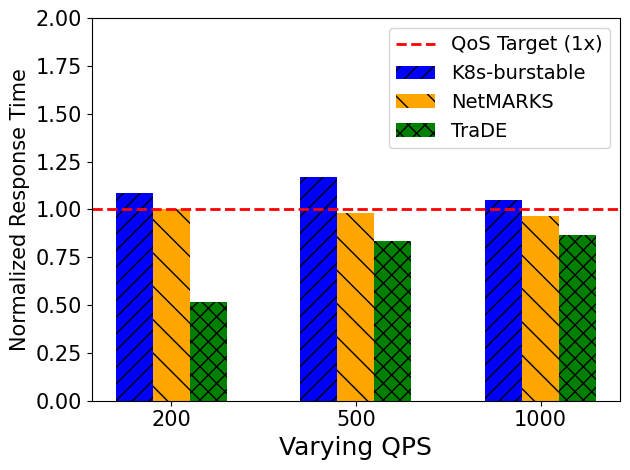}
        \label{fig:rt_user-timeline}
    }%
    \hfill
    \subfloat[\texttt{read-home-timeline}.]{
        \includegraphics[width=0.24\textwidth]{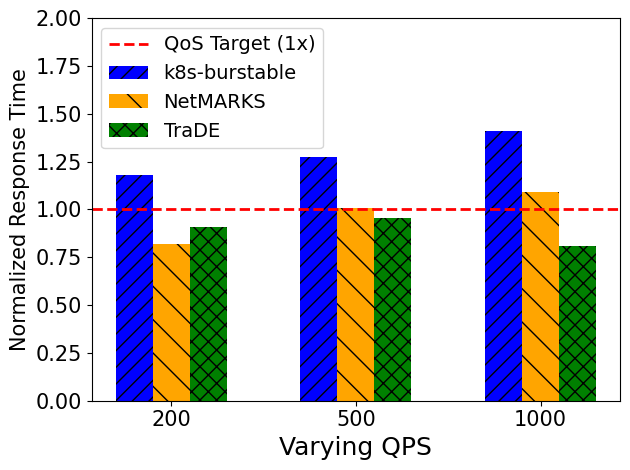}
        \label{fig:rt_home-timeline}
    }%
    \hfill
    \subfloat[Durations for execution time.]{
        \includegraphics[width=0.23\textwidth]{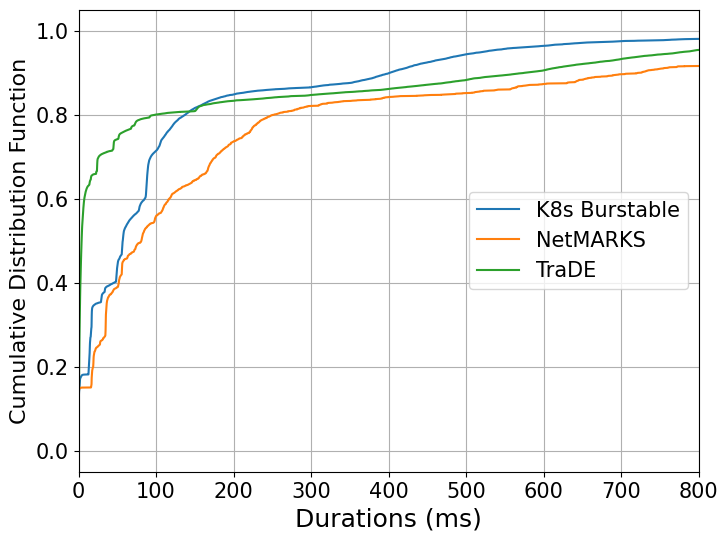}
        \label{fig:cdf_mix_duration}
    }
    \caption{In (a)(b)(c), different response time comparisons under varying QPS and request types. In (d), \textcolor{black}{the cumulative distribution function (CDF) figures of all triggered microservices execution time by mixed workload requests (with 6:2:2 ratio of \texttt{compose-post}, \texttt{read-user-timeline}, and \texttt{read-home-timeline} requests), showing TraDE has an overall reduced execution time, thereby leading to better end-to-end performance}.}
    \label{fig:social_net_RT}
\end{figure*}

\begin{figure*}[t]
    \centering
    \subfloat[\texttt{compose-post}.]{
        \includegraphics[width=0.23\textwidth]{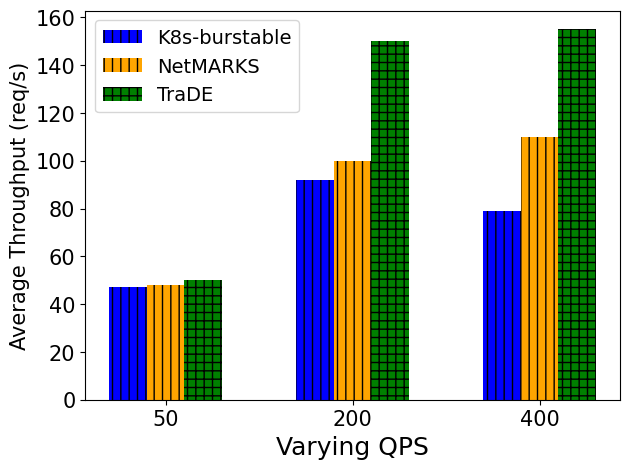}
        \label{fig:thput_compose-post}
    }%
    \hfill
    \subfloat[\texttt{read-user-timeline}.]{
        \includegraphics[width=0.23\textwidth]{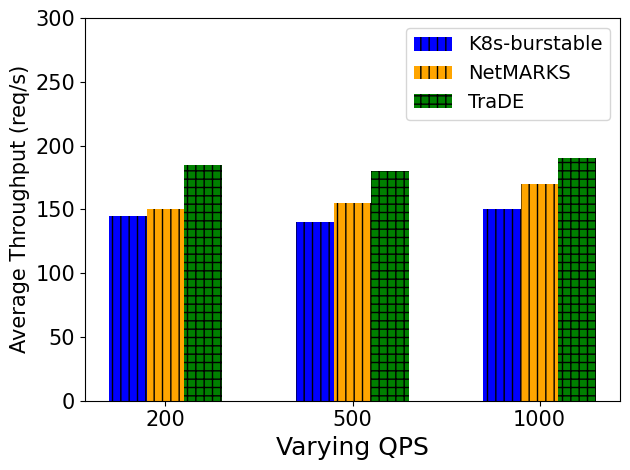}
        \label{fig:thput_user-timeline}
    }%
    \hfill
    \subfloat[\texttt{read-home-timeline}.]{
        \includegraphics[width=0.23\textwidth]{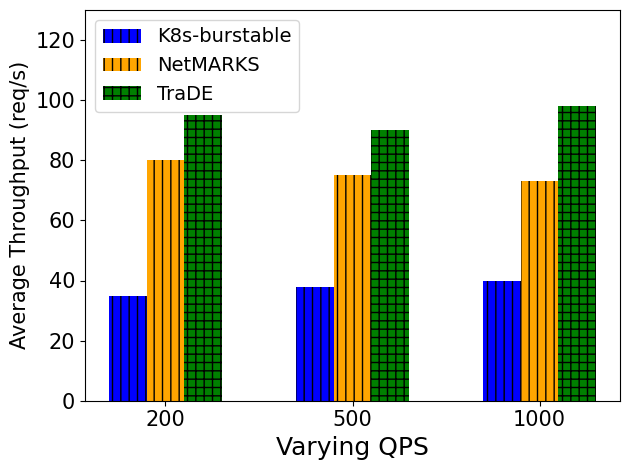}
        \label{fig:thput_home-timeline}
    }%
    \hfill
    \subfloat[Mixed requests (6:2:2).]{
        \includegraphics[width=0.23\textwidth]{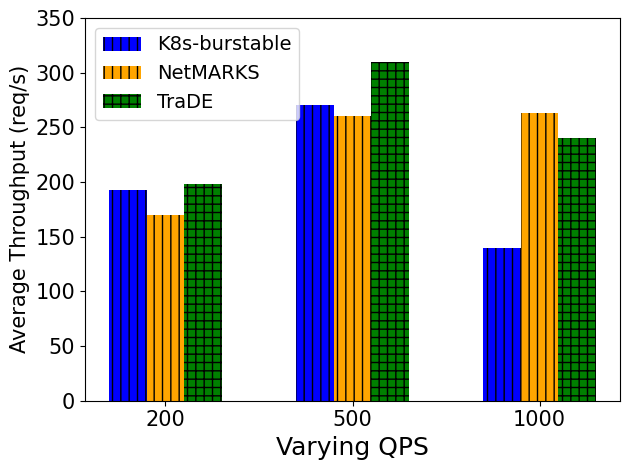}
        \label{fig:thput_mix622}
    }
    \caption{In (a)(b)(c), the evaluation comparisons of average throughput are shown under varying QPS and different request call-graphs. In (d), the throughput under mixed workload requests is displayed with the proportion of 6:2:2 for \texttt{compose-post}, \texttt{read-user-timeline}, and \texttt{read-home-timeline}, respectively.}
    \label{fig:social_net_Thrput}
\end{figure*}

\subsection{Workload Generator}
We adopted wrk2 \cite{wrk2} as the workload generator. As a modern HTTP benchmarking tool, wrk2 is capable of generating different types and proportions of workload requests for performance testing and measuring how well the cloud applications can handle varied traffic. Its ability to maintain a constant request rate makes it particularly useful for understanding the end-to-end performance of a server under controlled workload conditions.

\subsection{QoS Targets and Compared Methods}
\begin{figure*}[htbp]
  \centering
  \subfloat[QoS Target $<$ 250 ms\label{fig:QoS250}]{
    \includegraphics[width=.32\linewidth]{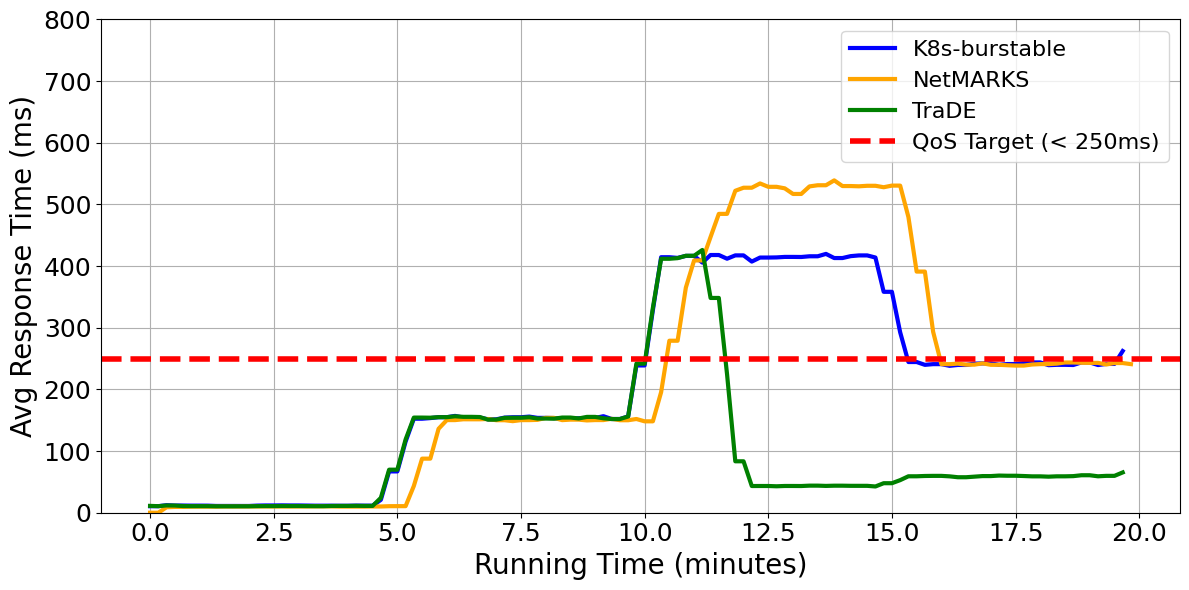}}
  \hfill
  \subfloat[QoS Target $<$ 300 ms\label{fig:QoS300}]{
    \includegraphics[width=.32\linewidth]{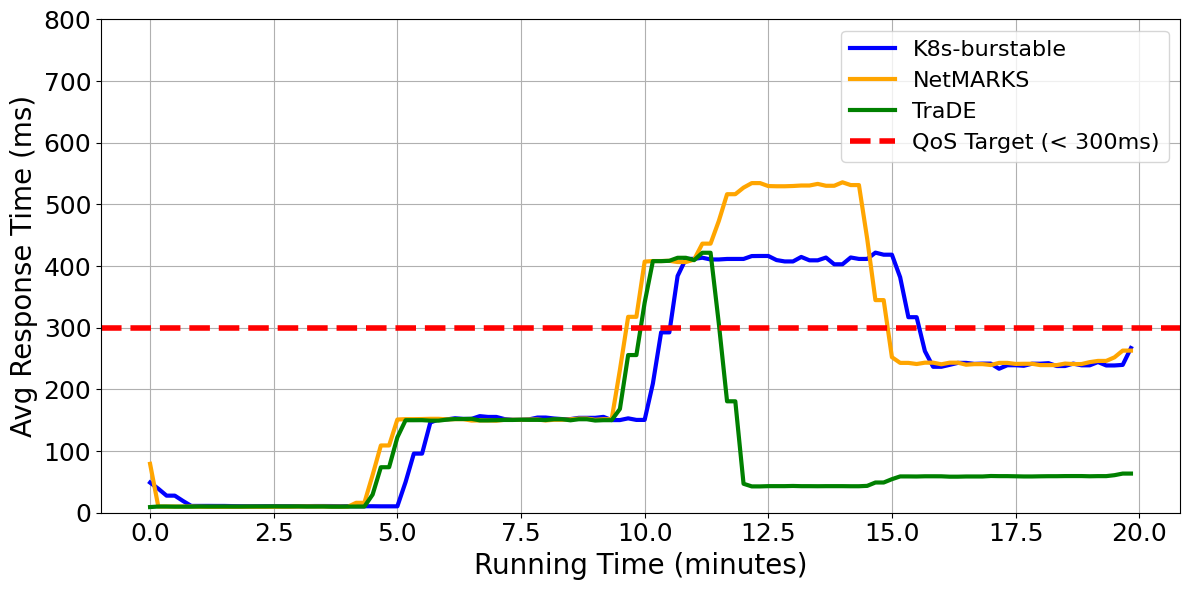}}
  \hfill
  \subfloat[QoS Target $<$ 350 ms\label{fig:QoS350}]{
    \includegraphics[width=.32\linewidth]{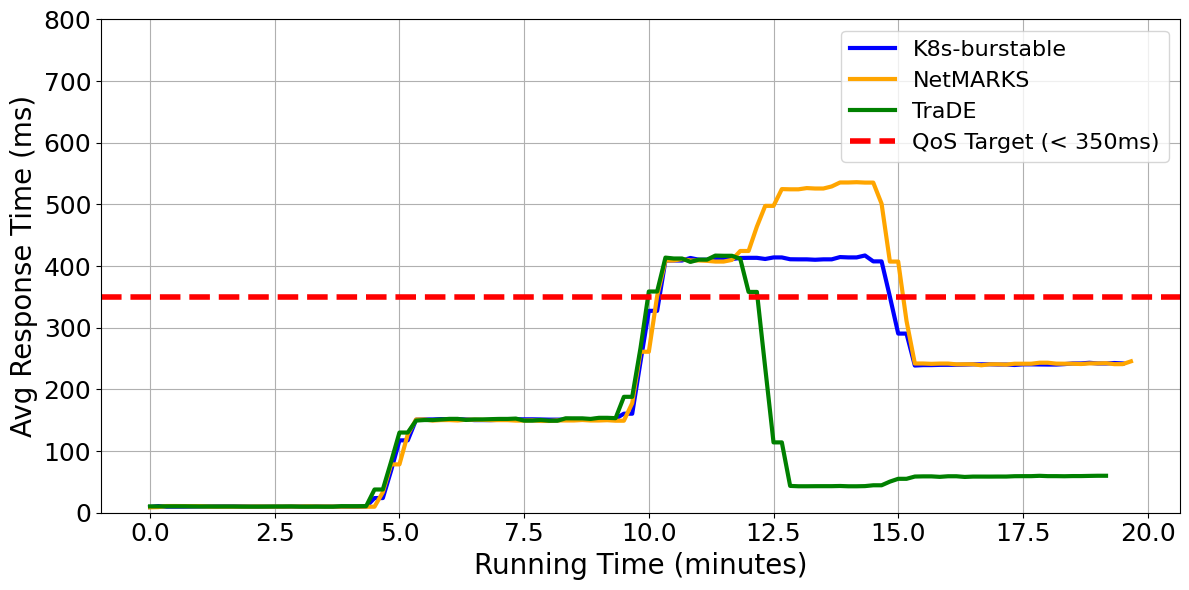}}

  \caption{\textcolor{black}{Under three different QoS targets and varying cross-node delays, the adaptive response of k8s-burstable, NetMARKS, and TraDE. In 20 minutes of the test time, 0 $\sim$ 5 mins, there are no injected cross-node delays; 5$\sim$10 mins, light cross-node delays are injected; 10$\sim$15 mins, heavy cross-node delays are injected; 15$\sim$20 mins, medium cross-node delays are injected. }}
  \label{fig:dynamicQoS}
\end{figure*}

{\color{black}
\subsubsection{QoS Target Determination (Trigger-Oriented)}
In our experiments, we use a fixed latency QoS target as the control-plane trigger for \texttt{TraDE}: every $\tau\!=\!30\,\text{s}$ the scheduler queries Prometheus over a sliding window $W$ (default $W\!=\!1\,\text{min}$) for Istio metrics in the target namespace with \texttt{response\_code}{=}\texttt{200}—specifically \texttt{istio\_request\_duration\_milliseconds\_sum} and \texttt{istio\_request\_duration\_milliseconds\_count} (via \texttt{rate} and \texttt{custom\_query\_range} functions)—and computes the windowed average response time $\bar L \!=\! \frac{\sum \texttt{sum\_values}}{\sum \texttt{count\_values}}$ (ms); if $\bar L \!>\! T$ with $T\!=\!300$\,ms (we also test $T\!\in\!\{250,300,350\}$\,ms in Fig. \ref{fig:dynamicQoS}), \texttt{TraDE} sets \textit{Trigger}{=}\texttt{True} and executes the rescheduling pipeline implemented in \texttt{TraDE} (traffic-graph construction, current placement extraction, and PGA-based remapping/migration); if the window has no data or the traffic count is zero, no trigger is raised.
}

\subsubsection{Compared Methods} To evaluate our proposed \texttt{TraDE} framework, we compared it with the default Kubernetes scheduling policy and the recent traffic-aware NetMARKS \cite{NetMARKS_InfoCOM21} which also targets network-aware scheduling for microservice applications.

\textbf{K8s default policy}: In Kubernetes, the default scheduling and Quality of Service (QoS) policies are designed to evenly distribute workloads across the cluster without further rescheduling policies even when microservice performance is violated.
QoS policies in the k8s cluster classify pods into three classes: \textit{Guaranteed}, \textit{Burstable}, and \textit{BestEffort}. \textit{Guaranteed} provides the highest priority ensuring pods always get the requested resources, \textit{Burstable} offers a flexible middle ground where pods have guaranteed minimum resources but can consume more if available, while \textit{BestEffort} has the lowest priority, where pods have no resource guarantees and can be preempted first during resource contention. In the evaluation experiments, we will use the default scheduling policy with \textit{Burstable} QoS class.


\begin{figure*}[t]
    \centering
    \subfloat[delay\_matrix1\_zero.]{
        \includegraphics[width=0.23\textwidth]{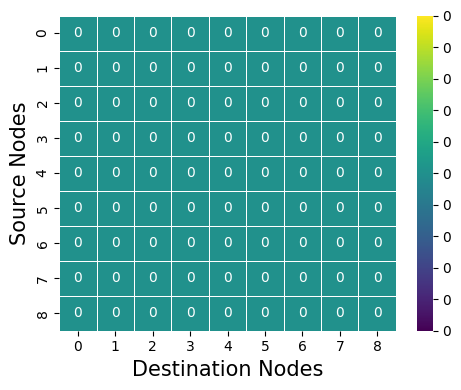}
        \label{fig:delay_matrix1_zero}
    }%
    \hfill
    \subfloat[delay\_matrix2\_light.]{
        \includegraphics[width=0.23\textwidth]{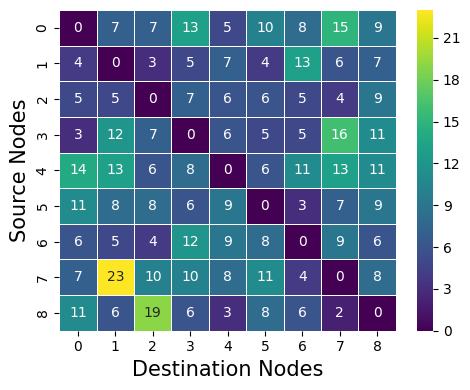}
        \label{fig:delay_matrix2_light}
    }%
    \hfill
    \subfloat[delay\_matrix3\_heavy.]{
        \includegraphics[width=0.23\textwidth]{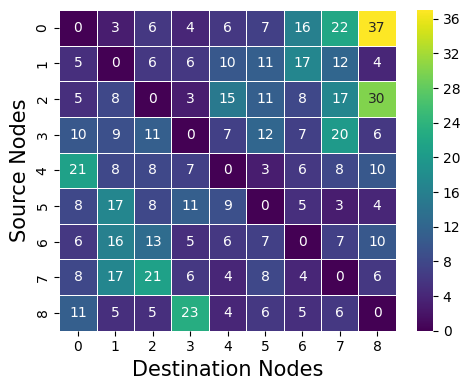}
        \label{fig:delay_matrix3_mid}
    }%
    \hfill
    \subfloat[delay\_matrix4\_medium.]{
        \includegraphics[width=0.23\textwidth]{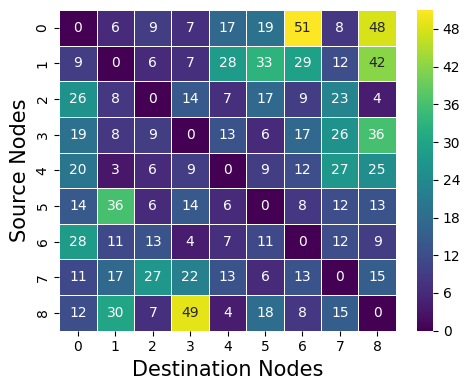}
        \label{fig:delay_matrix4_heavy}
    }
    \caption{\textcolor{black}{In (a)(b)(c)(d), dynamic delays are injected to the nine worker nodes.}}
    \label{fig:dynamic_delays}
\end{figure*}

\textbf{NetMARKS}:
The recent work NetMARKS \cite{NetMARKS_InfoCOM21} introduces a microservice pod scheduling scheme that leverages dynamic network metrics collected from the Istio Service Mesh. The main idea of NetMARKS is the proposed node scoring algorithm, which calculates node scores for a target pod by iteratively analyzing all pods running on each node and identifying those with traffic connections to the target pod. Each node's score is calculated based on the sum of traffic flows between the target pod and the selected pods on that node. The node with the highest score is then chosen to host the target pod.

With the implemented modules of \textit{Traffic Analyzer} in Section V, \textit{Dynamics Manager} in Section VI and \textit{PGA Mapper} in Section VII, we evaluated the proposed \texttt{TraDE} with benchmark microservice application and compared the end-to-end performance with default k8s QoS policy \cite{kubernetes} and NetMARKS \cite{NetMARKS_InfoCOM21}. In the evaluation experiments, we implemented the K8s QoS policy with \textit{Burstable} and the pod scheduling policy without any pre-set rules like affinity and taints. For NetMARKS \cite{NetMARKS_InfoCOM21}, we implemented the node scoring algorithm proposed by NetMARKS and adopted it for rescheduling the predefined target microservice pods experiencing QoS violations.

To ensure fair evaluations, three separate namespaces are created for three identical \texttt{social network} applications from \cite{deathStarBench_ASPLOS19}, each managed by a different method (i.e., K8s default, NetMARKS, and TraDE). Additionally, sustained workloads are sent concurrently to each of the three identical \texttt{social network} applications. This setup isolates the workloads for each method while maintaining the same cross-node delay settings for the three applications at the cluster level.

 We generated multiple workloads with different requests and varying QPS to evaluate the end-to-end performance of our proposed \texttt{TraDE} and existing methods. 

\begin{figure}[h]
    \centering
    \includegraphics[width=\linewidth]{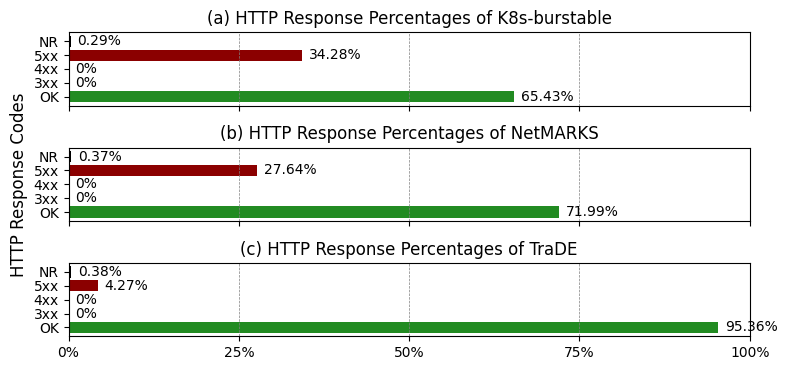}
    \caption{In each service mesh, HTTP response percentage distributions show the overall microservice application performance under different deployment methods. The higher the percentages of 'OK' responses, the higher the goodput ratio.}
    \label{fig:http_response_ratio}
\end{figure}

 \textbf{Response Time and Durations.}
For the \texttt{social network} benchmark application, we used the wrk2 tool to generate three request types: \texttt{compose-post}, \texttt{read-user-timeline}, and \texttt{read-home-timeline}. Each type shows distinct call graphs and traffic patterns across the application’s dependency graph. Fig. \ref{fig:social_net_RT} compares the average response times under various workloads, including changes in QPS and user requests. For these request types, \texttt{TraDE} consistently outperforms existing methods, meeting QoS targets across scenarios. Compared to NetMARKS \cite{NetMARKS_InfoCOM21}, \texttt{TraDE} achieves up to 12.3\% lower response times for \texttt{compose-post}, 48.3\% for \texttt{read-user-timeline}, and 25.8\% for \texttt{read-home-timeline} requests. Notably, \texttt{TraDE} adapts effectively under varied workloads, where K8s-burstable and NetMARKS do not always meet QoS.

Fig. \ref{fig:cdf_mix_duration} illustrates the CDF of execution times for all triggered microservices under a mixed workload (6:2:2 ratio of \texttt{compose-post}, \texttt{read-user-timeline}, and \texttt{read-home-timeline} requests). Here, \texttt{TraDE} demonstrates reduced microservice execution time, leading to lower response times and higher throughput.

\subsection{End-to-end Performance}

\textbf{Throughput and Goodput.} 
Throughput and goodput are key metrics for assessing end-to-end performance in dynamic environments. Throughput represents total data transmitted, including overhead, while goodput measures only the useful data successfully received at the application layer.


In terms of \textit{throughput}, we conducted experiments to measure the average throughput (requests/second) across different QPS and mixed workloads. As shown in Fig. \ref{fig:social_net_Thrput}, \texttt{TraDE} achieves higher throughput than NetMARKS \cite{NetMARKS_InfoCOM21}—up to 1.5x for \texttt{compose-post}, 1.2x for \texttt{read-user-timeline}, 1.4x for \texttt{read-home-timeline}, and 1.2x for mixed requests, indicating superior throughput in various scenarios.


For \textit{goodput}, we analyzed response types using Istio Service Mesh across separate deployments (K8s-burstable, NetMARKS, and TraDE) for isolation. Fig. \ref{fig:http_response_ratio} shows \texttt{TraDE} achieves a 95.36\% success rate, outperforming NetMARKS (71.99\%) and K8s-burstable (65.43\%). These results show that \texttt{TraDE} surpasses existing methods in both throughput and goodput across dynamic workloads.

\subsection{Adaptive Performance Under Changing Delays}

To assess the adaptive capability of the proposed \texttt{TraDE} framework, we evaluated its performance under fluctuating cross-node communication delays. \textcolor{black}{Specifically, in Fig. \ref{fig:dynamic_delays}, four different cross-node delays were injected to the cluster nodes every five minutes.
As shown in Fig. \ref{fig:dynamicQoS}, \texttt{TraDE} effectively responds to these changing delays by adaptively redeploying microservice instances to meet QoS targets once detecting QoS violations. Comparing with the existing two methods, it is clear to observe that TraDE can consistently maintain response times within the QoS targets (i.e., $< 250 ms, 300 ms, <350 ms$) throughout the remaining runtime, while the other two methods (K8s-burstable and NetMARKS) failed to meet these QoS targets.
}


\section{Conclusions and Future Work}

In this work, we designed a traffic and network-aware framework, \texttt{TraDE}, to address the challenges of QoS violations in containerized microservices running in dynamic computing environments. Our framework primarily consists of three components: a traffic stress analyzer, a network dynamics manager, and an efficient service node mapper. We evaluated our proposed \texttt{TraDE} against existing solutions and demonstrated that our framework effectively meets the QoS targets under various dynamic conditions, outperforming the existing method NetMARKS by reducing response time by up to 48.3\%, improving throughput by up to 1.4x and showing robust adaptiveness under sustained workloads.

As part of future work, we plan to enhance our framework to support resource auto-scaling for microservice-based GPT (Generative Pre-trained Transformer) applications at the cloud-edge continuum, which demands stricter end-to-end performance, and they are sensitive to dynamics like cross-node delays.

\textbf{Software}: \texttt{TraDE} software code as open source is available at \href{https://github.com/Cloudslab/TraDE}{https://github.com/Cloudslab/TraDE}



\bibliographystyle{ieeetr}
\bibliography{reference}

\vspace{-30pt}
\begin{IEEEbiography}[{
\includegraphics[width=1in,height=1.25in,clip,keepaspectratio]{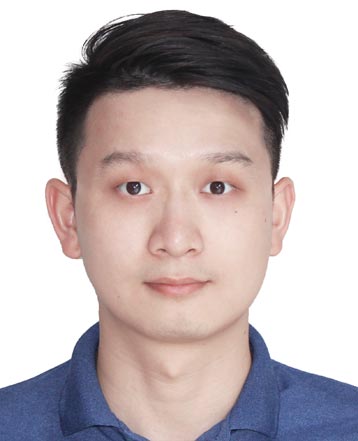}
}]
{Ming Chen} is a PhD candidate in the School of Computing and Information Systems (CIS) at the University of Melbourne, Australia. His main research interests focus on resource scheduling, networking, and cloud computing. 
\end{IEEEbiography}


\vspace{-50pt}
\begin{IEEEbiography}[{
\includegraphics[width=1in,height=1.25in,clip,keepaspectratio]{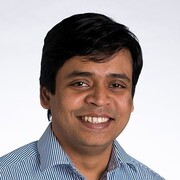}
}]
{Muhammed Tawfiqul Islam} is a lecturer in the School of Computing and Information Systems (CIS) at the University of Melbourne, Australia. He completed his PhD from the School of Computing and Information Systems (CIS) at the University of Melbourne, Australia in 2021. His research interests include resource management, cloud computing, and big data.
\end{IEEEbiography}

\vspace{-50pt}
\begin{IEEEbiography}[{
\includegraphics[width=1in,height=1.25in,clip,keepaspectratio]{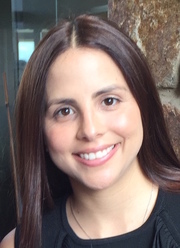}
}]
{Maria Rodriguez Read} is a lecturer in the School of Computing and Information Systems at the University of Melbourne, Australia. Her research focuses on distributed and parallel systems, particularly optimizing support for containerized and cloud-native applications to improve scalability, robustness, and cost-efficiency in cloud deployments.
\end{IEEEbiography}

\vspace{-50pt}
\begin{IEEEbiography}[{
\includegraphics[width=1in,height=1.25in,clip,keepaspectratio]{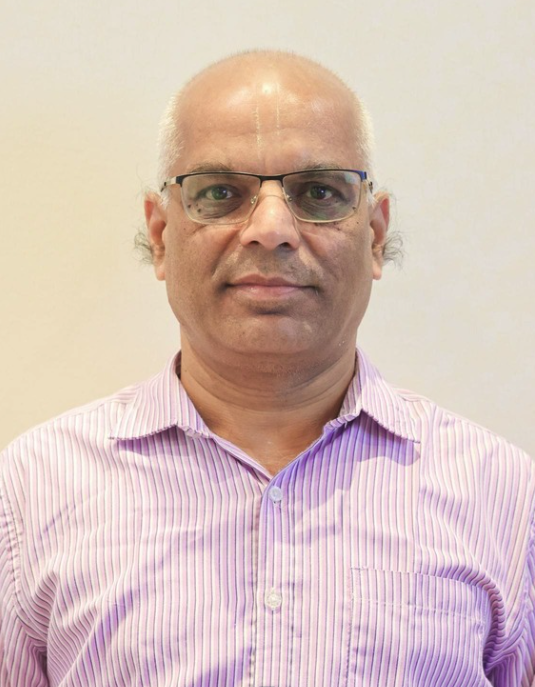}
}]
{Rajkumar Buyya} (Fellow of IEEE \& Academia Europea) is a Redmond Barry distinguished professor and director of the Cloud Computing and Distributed Systems (CLOUDS) Laboratory, University of Melbourne, Australia. He has authored over 800 publications and seven textbooks. He is one of the highly cited authors in computer science and software engineering worldwide (h-index=175, g-index=384, 164,100+ citations).
\end{IEEEbiography}

\end{document}